\documentclass[10pt,conference]{IEEEtran}

\usepackage{booktabs} 
\usepackage{multirow}

\usepackage{cite}
\usepackage{listings}
\usepackage{amsmath,amssymb,amsfonts}
\usepackage{algorithmic}
\usepackage{graphicx}
\usepackage{textcomp}
\usepackage{multirow}
\usepackage{caption}
\usepackage{amsmath}
\usepackage{amssymb}
\usepackage{multirow} 
\usepackage{xcolor}

\usepackage{color}
\usepackage{booktabs}
\usepackage{colortbl}

\usepackage{etoolbox}
\usepackage{array}
\usepackage{float}
\usepackage{diagbox}
\usepackage{xspace}
\usepackage[british,english]{babel}
\usepackage{graphicx}
\usepackage{subfig}
\usepackage{balance}
\usepackage{url}
\usepackage[ruled,linesnumbered]{algorithm2e}
\usepackage[normalem]{ulem}
\usepackage{soul}

\newcommand\dl[1]{\textcolor{blue}{\sout{#1}}}

\DeclareMathOperator*{\argmin}{arg\,min}

\newcommand{\circled}[2][]{\tikz[baseline=(char.base)]
    {\node[shape = circle, draw, inner sep = 1pt]
    (char) {\phantom{\ifblank{#1}{#2}{#1}}};%
    \node at (char.center) {\makebox[0pt][c]{#2}};}}
\robustify{\circled}

\definecolor{Gray}{gray}{0.9}

\newcommand{\PCA}{\texttt{PCA}\xspace}

\newcommand{\PCs}{\texttt{PCs}\xspace}

\newcommand\cparagraph[1]{\vspace{2.5mm} \noindent \textbf{#1.}\xspace}

\renewcommand\dl[1]{} 
\renewcommand\hl[1]{#1} 

\IEEEoverridecommandlockouts
\begin{document}

\title{Using Machine Learning to Optimize Web Interactions on Heterogeneous Mobile Systems}

\author{\IEEEauthorblockN{Lu Yuan\IEEEauthorrefmark{3}, Jie Ren\IEEEauthorrefmark{2}, Ling Gao\IEEEauthorrefmark{3}, Zhanyong Tang\IEEEauthorrefmark{3},
         Zheng Wang\IEEEauthorrefmark{1}}

 \IEEEauthorblockA{\IEEEauthorrefmark{3} Northwest University, China, \\ \IEEEauthorrefmark{2} Shaanxi Normal University, China, \\
 \IEEEauthorrefmark{1} University of Leeds, United Kingdom}
 }

\maketitle

\begin{abstract}
The web has become a ubiquitous application development platform for mobile systems. Yet, web access on mobile devices remains an
energy-hungry activity. Prior work in the field mainly focuses on the initial page loading stage, but fails to exploit the opportunities
for energy-efficiency optimization
 while the user is interacting with a loaded page. This paper presents a novel approach for performing energy optimization for
interactive mobile web browsing. At the heart of our approach is a set of machine learning models, which estimate \emph{at runtime} the
frames per second for a given user interaction input by running the computation-intensive web render engine on a specific processor core
under a given clock speed. We use the learned predictive models as a utility function to quickly search for the optimal processor setting
to carefully trade responsive time for reduced energy  consumption. We integrate our techniques to the open-source Chromium browser and
apply it to two representative mobile user events: scrolling and pinching (i.e., zoom in and out). We evaluate the developed system on
the landing pages of the top-100 hottest websites and two big.LITTLE heterogeneous mobile platforms. Our extensive experiments show that
the proposed approach reduces the system-wide energy consumption by over 36\% on average and up to 70\%. This translates to an over 17\%
improvement on energy-efficiency over a state-of-the-art event-based web browser scheduler, but with significantly fewer violations on
the quality of service.
\end{abstract}

\begin{IEEEkeywords}
Interactive Mobile Web Browsing, Machine Learning, Energy Optimization
\end{IEEEkeywords}

\section{Introduction}
In recent years, portable mobile devices like smartphones and tablets have become the dominant personal computing
platform~\cite{marketreport}. Concurrent to this mobile computing evolution is the wide adoption of web technology as a development
platform for many mobile applications like web browsing, social networking, news reading and online banking. Indeed, the web has become a
major information portal for mobile systems and accounted for two-thirds of the mobile traffics in the US~\cite{mobiletraffic}.

Energy and performance optimization for mobile web browsing is an open problem. \dl{Mobile users want their devices to appear fast and
responsive but at the same time to have long-lasting batteries. Achieving both at once is difficult.} Like many other mobile applications,
 the performance-energy \hl{trade-off} is a critical issue for interactive mobile web browsing, because users expect a degree of
responsiveness when browsing a webpage, but also want low energy consumption when interacting with their battery-powered devices.

Recently, efforts have been made to improve the energy efficiency for mobile web browsing by focusing on the initial page loading
phase~\cite{ren2017optimise,Ren:2018:PNW:3281411.3281422}. These prior approaches exploit the performance-energy elasticity provided by the
heterogeneous multi-core hardware design to trade page loading time for lowered power consumption. Although impressive results were shown,
such approaches do not consider the impact of user interactions after the initial page loading stage. However, as we will show later in the
paper, user interactions can account for a large portion of power consumption for mobile web browsing and thus cannot be ignored for energy
optimization.

Some of the more recent studies like PES~\cite{pes} and eBrowser~\cite{xu2018ebrowser} have attempted to address the energy optimization
problem for interactive mobile web browsing. While promising, they have critical drawbacks. Specifically, PES employs an analytical model
to choose the operating frequency of the processor to reduce energy consumption. Developing an effective analytical model requires deep
knowledge of the underlying hardware and the application domains~\cite{wang2018machine}. As a result, PES offers a poor hardware
portability because tuning a model for a new hardware platform could involve significant overhead. While eBrowser avoids the pitfall of
using an analytical model, it performs energy optimization by simply dropping some of the user events (which is not ideal as it can miss
some important user inputs) and does not explore the rich optimization space offered by the increasingly popular heterogeneous multi-core
design. Since heterogeneous multi-cores like ARM big.LITTLE~\cite{biglittle} have become the \emph{de facto} design choice for mobile
systems, eBrowser leaves much room for improvement.

In this work, we aim to close the gap of energy optimization for interactive mobile web browsing. Our goal is to design an adaptive scheme
to unlock the potential of heterogeneous multi-cores to better \hl{perform} energy optimization for interactive mobile web browsing. Our key
insight is that a slight delay in responding to a user input might be imperceivable to or acceptable by the
user~\cite{seeker2014measuring}, but this provides chances for reducing the energy consumption of a mobile system -- by running the
computation-intensive rendering process on right processor with an appropriate (but not necessarily the highest) clock speed. In other
words, if we can carefully trade the processing time to provide ``just good enough" responsiveness, we can then reduce the energy
consumption of the entire system without significantly compromising the quality of service (QoS).

While intuitive, translating this idea to build a practical system is not trivial. The key challenge here is how to develop an effective
scheme that can be portable across different mobile platforms. Given the diversity of today's mobile devices and the constantly evolving
nature of hardware design -- where the number of processor cores and capabilities of a mobile device are likely to change from one
generation to the other -- it is important to make sure whatever strategies we developed today can be easily ported to and deliver good
performance on a new hardware architecture tomorrow.

We address the portability issue by employing machine learning to automatically learn how to best configure the underlying heterogeneous
multi-core hardware for a given web page to meet the QoS requirement for a specific user and event. Our machine-learning models are
automatically built offline using training web pages, and the learned models can be applied to any \emph{new, unseen} web content. This
automatic learning-based scheme removes the need for manually rewriting an analytical model every time the hardware has changed. By
reducing the expert involvement, our approach thus reduces the cost of model construction and offers a better  generalization ability and
performance portability.


To evaluate our approach, we have developed a working prototype\footnote{We stress that our techniques can be applied to not only web
browsers but also many mobile applications that use web rendering technology.} based on Chromium~\cite{Chromium} -- the open-source
backbone of \hl{many mainstream web browsers including Google Chrome  and Microsoft Edge-for-ARM64}. We apply the developed system to the
landing pages of the top-100 popular websites ranked by \texttt{\url{www.alexa.com}}~\cite{alexa} and consider two representative mobile
interactions: scrolling and pinching (i.e., zoom in and out). We evaluate our work on two distinct heterogeneous mobile platforms: Odroid
Xu3 and Jetson TX2. Experimental results show that our approach consistently outperforms the state-of-the-art on all evaluation metrics, by
delivering over 17\% more energy reduction but with fewer QoS violations.

In summary, the key contribution of this paper is a new approach for optimizing interactive mobile web browsing on heterogeneous multi-core
mobile platforms. Compared to existing solutions, our approach has the benefits of being low-cost for model construction and portable
across hardware architectures. We show that these benefits do not come at the cost of performance penalties. By contrast, our approach
delivers consistently better performance over the state-of-the-art across web content and evaluation platforms.

\section{Related Work}

Our work lies at the interaction of the following five areas but qualitatively differs from prior works within each area.

\subsection{Energy optimization} Energy and power optimization for embedded and mobile systems is an intensely studied field. There is a
wide range of activities on exploiting compiler-based code optimization~\cite{wang2015automatic, moran2018overcoming}, runtime task
scheduling~\cite{ Taylor:2017:AOO:3078633.3081040,taylor2018adaptive}, or a combination of both~\cite{ wang2018machine} to optimize
different workloads for energy efficiency. Other relevant work in web browsing optimization exploits application knowledge to batch network
communications ~\cite{6848020, li2016automated}, and parallel downloading~\cite{ sehati2017energy}, which primarily target the initial page
loading phase. Our work is complementary to prior works by targeting the low-level optimization, and we do so by utilizing the hardware
configuration knobs to perform energy optimization during the interacting phase.

\subsection{Optimization for Web Access}
Our work is closely related to research on optimizing web browsing. Prior works have shown that by carefully choosing the processor
frequency, one can reduce the energy consumption for the initial page loading phase~\cite{zhu2013high,ren2017optimise,
Ren:2018:PNW:3281411.3281422}. PES~\cite{pes}  and eBrowser~\cite{xu2018ebrowser} are most closely related to our work.  PES employs an
analytical model to choose the optimal processor frequency and does not consider the impact of web content to the responsive time.
Developing an effective analytical model requires insight knowledge of the underlying hardware~\cite{wang2018machine}, which makes it
difficult for the approach to be adopted by new hardware architectures. Our work avoids this pitfall by using machine learning to
automatically learn a portable approach for how to best optimize for interactive events. Furthermore, we show that web content can have a
significant impact on the processor response time, and cannot be ignored. For this reason, our approach explicitly captures and models the
impact of web content. Like eBrowser, our approach also trades responsive time for energy efficiency. \dl{Our work advances eBroswer by
exploiting the heterogeneous mobile architecture and processor frequency settings to reduce energy consumption. By exploring a larger
optimization space, we achieve better energy efficiency over eBrowser.} \hl{Unlike eBrowser, we perform energy optimizations through
dynamic voltage and frequency scaling (DVFS) and task scheduling, while eBrowser achieves this by dropping events via a simple sleeping
mechanism. In other words, eBrowser does not utilize the hardware configuration knobs provided by a big.LITTLE architecture design. As a
result, our work is complementary to the event dropping mechanism of eBrowser. } A recent work~\cite{Peters:2018:PWB:3205289.3205293}
proposed a phase-aware power management scheme to control the processor power state of different web browser phases like loading and
touching. This approach considers a fixed response latency threshold for a given phase. Unlike ~\cite{Peters:2018:PWB:3205289.3205293}, we
offer a more flexible, personalized approach by considering the impact of web content on the user perceive latency and the diverse
expectations across different users.

Our work builds upon and directly benefits past foundations on web workload
characterization~\cite{thiagarajan2012killed,cao2017deconstructing}. Other studies exploit the interplay between the web server and browser
client to improve rendering speed and user experience~\cite{butkiewicz2015klotski, netravali2018prophecy}, or reconstruct the web browser
architecture~\cite{Qian:2014:CRU:2594368.2594372, bui2015rethinking}. These works are thus orthogonal to our approach.

\subsection{Task Scheduling}
As heterogeneous multi-cores are becoming the norm of computing systems, how to effectively schedule application tasks on such
architectures have attracted intensive attention. There is considerable work on designing better heuristics or models to schedule
application tasks for performance and energy optimization~\cite{Augonnet:2011:SUP:1951453.1951454, mittal2015survey, chronaki2016task,
castillo2018architectural}. Our work targets an important domain of mobile web browsing. It builds upon these past results to develop a
novel approach to exploit the characteristics of application workloads and hardware to better optimizing interactive mobile web browsing.
The main advantage of our machine learning based approach over a hand-crafted model or heuristic is the better portability - machine
learning enables one to automatically build a model for a new hardware design to adapt to the change of hardware.

\subsection{Machine Learning for System Optimization}

Machine learning has quickly emerged as a powerful design methodology for systems optimization. Prior works have demonstrated the success
of applying machine learning for a wide range of systems optimization tasks, including modeling personal preference on
wearables~\cite{krause2005context}, human activity recognition ~\cite{plotz2018deep,Zhang:2018:CTC:3241539.3241570}, code
optimization~\cite{wang2014integrating,Tournavitis:2009:THA:1542476.1542496,Wang:2009:MPM:1504176.1504189,wang2010partitioning,grewe2013portable,wang2013using,DBLP:journals/taco/WangGO14,
ogilvie2014fast,cummins2017end,ogilvie2017minimizing,spmv,ipdpsz18,ijpp18}, task
scheduling~\cite{grewe2011workload,emani2013smart,grewe2013opencl}, task
scheduling~\cite{grewe2011workload,emani2013smart,grewe2013opencl}, processor resource allocation~\cite{wen2014smart}, and many others.
\hl{These prior works provide strong evidence, showing that machine learning is a rigorous methodology for searching and extracting
application knowledge that can be transferred and reused in unseen settings for systems optimization. } In this work, we employ machine
learning techniques to develop an automatic and portable approach to optimize interactive mobile web browsing for energy efficiency. We
want to highlight that our work does not seek to advance the machine learning algorithm itself; instead, it exploits and applies a
well-established method of statistically reasoning  to tackle an important systems optimization problem, in a way that has not been
attempted.

\subsection{Heterogeneous Mobile Multi-cores}
\hl{Heterogeneous asymmetric chip-multiprocessor (ACMP) architecture is now commonplace on mobile
systems}~\cite{lo2015prediction,gaudette2016improving, Mishra:2018:CLC:3173162.3173184}. \hl{Recent examples in this space include the ARM
big.LITTLE architecture}~\cite{Seo2015Big}. \hl{Such design is adopted by major mobile hardware vendors including Qualcomm, Samsung, Huawei
and Apple.  The big.LITTLE architecture incorporates a performance-tuned high-end processor and an energy-tuned low-end processor, where
each processor provides a range of frequency settings to trade performance for power consumption. By providing a wide array of latency and
energy tradeoffs, this approach is claimed to deliver a 3x energy saving over the typical daily use profile of a smartphone with no loss of
perceived performance}~\cite{arm}. \hl{However, the hardware potential can only be unlocked by the application software. The challenge for
software optimization is that while the energy benefits of choosing the right heterogeneous core may be large, but mistakes can seriously
hurt the user experience}~\cite{lo2015prediction}. \hl{This paper proposes a software-based approach to exploit the optimization space
opened up by a heterogeneous multi-core design, specifically targeting interactive mobile web browsing -- an area that is largely
unexplored to date. Rather than developing a hand-crafted approach that requires expert insight into the relative costs and idiosyncrasies
of a particular heterogeneous multi-core, we develop an automatic technique that is independent of a particular platform.}

\section{Background \label{sec:background}}

\subsection{Problem Scope}

 Mobile web browsing includes two distinct phases~\cite{Peters:2018:PWB:3205289.3205293} for initial page loading and
responding to user inputs. During the page loading phase, web content will be fetched and parsed to construct a Document Object Model (DOM)
tree for rendering. Because a web page often cannot fit into a single screen view on a smartphone, only the currently visible area will be
rendered by the browser at a given moment. In the user interaction phase, the web browser responds to user events (e.g., scrolling) to
render and update the visible viewport accordingly. In this paper, we solely focus on the later interacting phase. We consider two typical
mobile interactive events: scrolling and pinching, but our approach can be applied to other user gestures too.

\subsection{Hardware architecture} Our work targets the ARM big.LITTLE heterogeneous multi-core design, the \emph{de facto} hardware architecture for modern mobile
systems. The big.LITTLE integrates an energy-tuned CPU processor (little) with a faster but more power-hungry processor (big). Such
heterogeneous design gives the flexibility for the software to choose a processor core to run a given task depending on the energy and time
constraints.

Specifically, our work is evaluated on two representative ARM big.LITTLE implementations: Odroid Xu3 and Jetson TX2, which integrate two
different generations of the big.LITTLE architecture. \dl{Having two distinct platforms allows us to evaluate the performance portability
and generalization ability of our approach.} We provide further details of our hardware platforms when describing our experimental setup in
Section~\ref{sec:setup}.

\section{Motivation of The Work}
\label{sec:motivation} To show the need for interactive mobile web browsing optimizations, we measured the energy spent by the Chromium
browser for responding to scrolling events when processing the landing page of the top-100 hottest websites ranked by \texttt{alexa.com}.
We automatically generate
 scrolling events to ensure all content of each page is shown on the screen, but ensure that each scrolling session only leads to a full-screen update. In this
experiment, we use an Jetson TX2 mobile development board\dl{In this
experiment, we use an Odroid Xu3 mobile development board that implements the hardware architecture used by the Samsung Galaxy S5
smartphone. Odroid Xu3}, \hl{which} integrates a  Cortex-A57 (big CPU cluster) and a Denver2 (little CPU cluster) CPUs, and a NVIDIA Pascal
GPU.

\begin{figure}[!t]
	\centering
    \vspace{-2mm}
	\subfloat[][Energy consumption]{\includegraphics[width=0.24\textwidth]{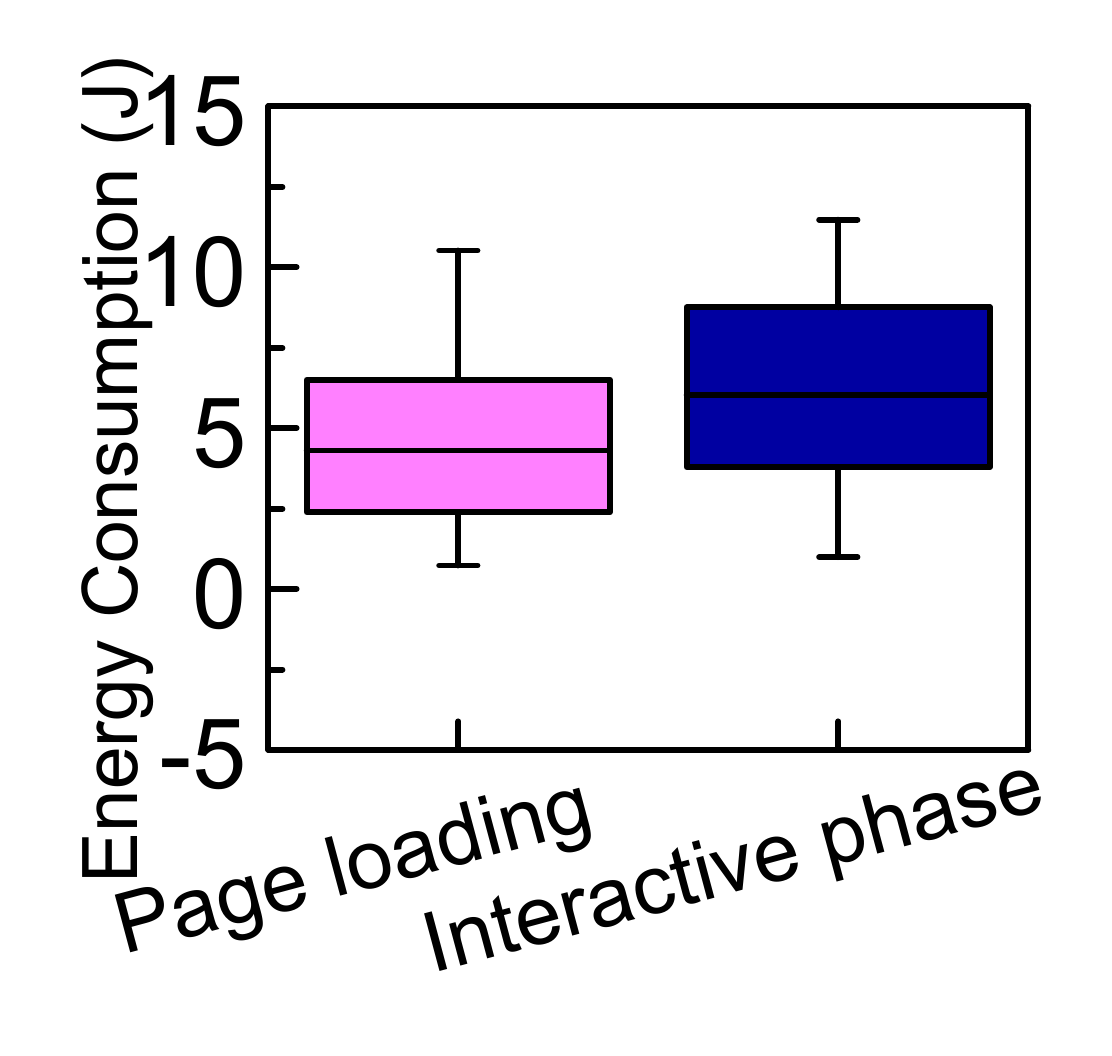}}
    \hfill
    \subfloat[][Time]{\includegraphics[width=0.24\textwidth]{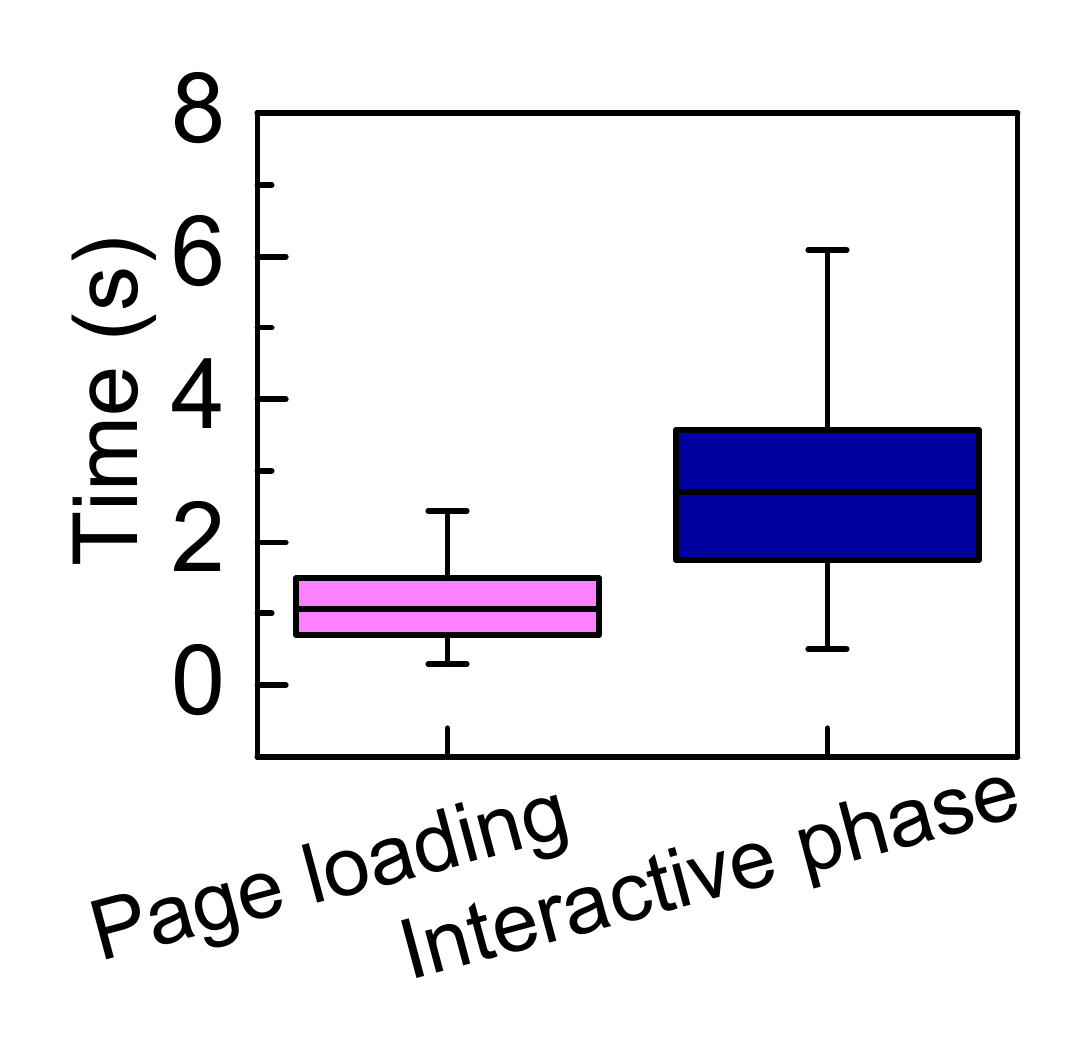}}

    \caption{Energy consumption (a) and time (b) for loading 100 webpages and responding to scrolling in a WiFi network. The min-max bar shows the variance across webpages.
    User interactions often lead to longer running and more energy consumption over the initial page loading phase, which thus cannot be ignored
    for energy optimization.
    }
    \label{fig:100web}
\end{figure}

Figure~\ref{fig:100web} compares the average energy and time spent during the loading and the interacting phases when browsing a webpage.
The min-max bars of the diagram show the variance across different pages. For each page, we profile the time and energy multiple times
until the 95\% confidence interval is within a range of 5\% variances to ensure the measurement is statistically sound. Note that our
measurement excludes time and energy during idle time.  Interactions on average are 94\% (up to 3x) longer and consume 44\% (up to 3.5x)
more energy than the initial page loading phase. This huge disparity between energy consumption and processing time suggests that previous
work which only focuses on the initial page loading phase would miss a massive opportunity for energy optimization.

\begin{figure}[!t]
	\centering
	\subfloat[][\hl{FPS}]{\includegraphics[width=0.24\textwidth]{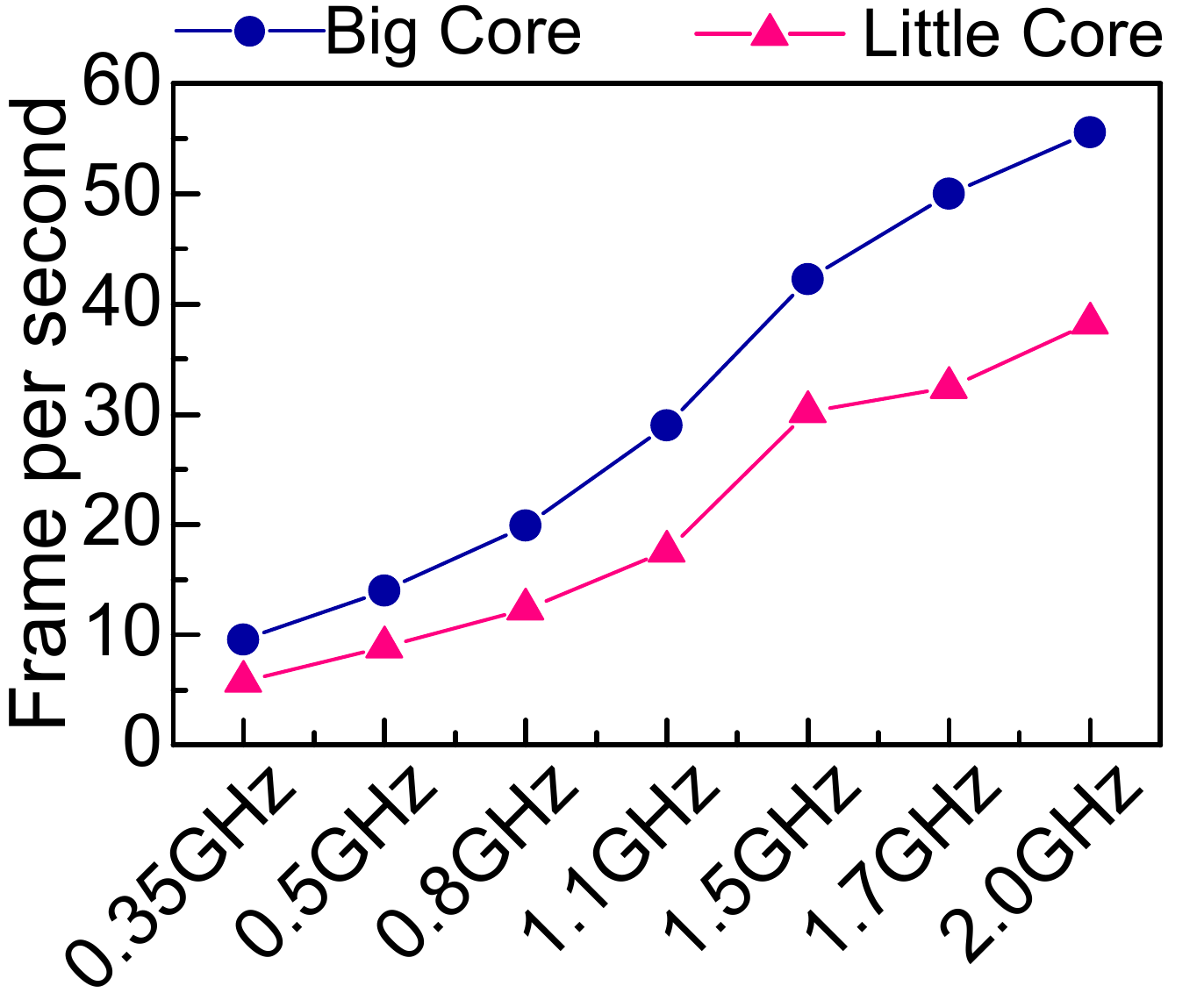}}
    \hspace{-5mm}
    \hfill
        \hspace{-5mm}
    \subfloat[][\hl{Energy consumption}]{\includegraphics[width=0.24\textwidth]{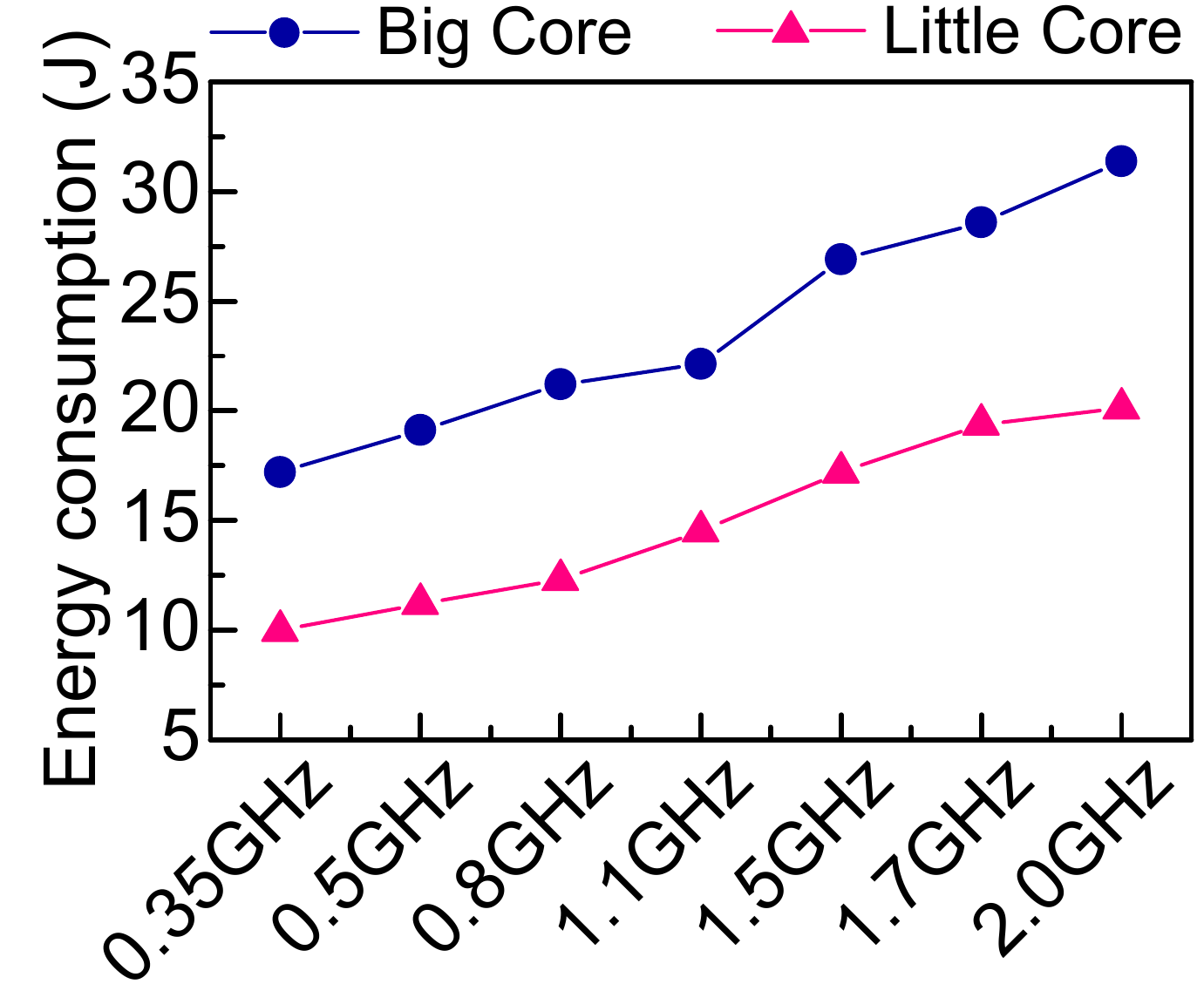}}
    \vspace{-2mm}
    \caption{FPS (a) and Averaged energy consumption (b) during the interactive phase when browsing the landing page of \texttt{cnn.com} on Jetson TX2. A
    bar shows the measured FPS and energy when running the computation-intensive render process on the big or little cluster with a specific processor clock frequency.}
    \label{fig:cnn}
\end{figure}

\dl{This work aims to close this gap by performing energy optimization during the interacting phase. Our key insight is that we can reduce
the power consumption by responding to user input using a lower CPU clock frequency, and a slight delay in screen updating might be
tolerable by the user.} \hl{This work aims to trade response time for reduced energy consumption.} To elaborate on our point, consider now
Figure~\ref{fig:cnn} which shows the \hl{energy consumption} and frames per second (FPS) under different CPU clock frequencies for the
landing page of \texttt{cnn.com} on Jetson TX2. In this example, we map the computation-intensive render process to run on a big or a little CPU cluster
under different clock frequencies, and the rest browser processes to run on the other cluster. \dl{Running the render process on the big
CPU cluster with the highest clock speed gives an FPS of 60.}\hl{Running the render process on the big cluster always gives a higher FPS
over the little counterpart under the same processor clock frequency. However, it is not always energy-efficient to use the big processor
cluster.} Our user study presented in Section~\ref{sec:qqos} shows that the FPS strongly correlates to the minimum acceptable
responsiveness of a user, but a typical user would be unable to tell the difference between 30 FPS and a higher screen update rate for web
browsing. This finding is in line with the observation presented in a prior study on mobile user experience~\cite{seeker2014measuring}.
This suggests that \hl{we can run the render process on the little cluster with 1.5 GHz instead of the big cluster or a higher processor
frequency.} \hl{Such a setting} already gives a ``good enough" FPS of 31 for this example \hl{but uses 36\% less energy compared to running
the render process on the big cluster with the same frequency.}

The question here is that ``what is the optimal\footnote{In this work, the optimal processor setting is a CPU cluster (little or big in our
case) with a specific clock frequency for running the render process, which gives the largest energy saving but also delivers the minimum
acceptable FPS for a given user.} processor setting to use?". The right answer depends on which CPU cluster (big or little) we choose to
use and at what frequency the CPU cluster will operate on. The choice also depends on the incoming user event rate and the web content,
because they determine how long it will take to update a screen view. Unfortunately, choosing the right processor setting is not trivial as
an inappropriate setting can lead to either an unacceptable FPS (spoiling the QoS) or unnecessarily higher energy consumption (wasting
battery life). In the next section, we will describe how to develop an adaptive scheme based on machine learning to maximize energy
reduction without necessarily compromising the QoS.


%

\section{Our Approach \label{sec:modelling}}
\subsection{Overview}
\dl{This work aims to reduce energy consumption during interactive mobile web browsing without compromising the QoS. We achieve this by
dynamically choosing the optimal CPU cluster and processor frequency to run the computation-intensive render process.}

The core of our approach is a set of regression models \hl{trained offline}, each is tuned for a specific user event. \dl{Our models are trained offline using
training data.} The trained models are then used to make predictions for \emph{new}, \emph{unseen} webpages. The model estimates the FPS
under a specific processor frequency, by taking into consideration the web content to be rendered, and the incoming user event and event
rate. The predictive model is used as a utility function to quickly search for the optimal processor setting. Predictions are made based on
a set of numerical values, or \emph{feature values}, described in Section~\ref{sec:features}.

Some prior works on optimizing the webpage loading follow a constrained based approach to use a classifier to choose from a set of
processor frequencies~\cite{ren2017optimise,Ren:2018:PNW:3281411.3281422}. However, these methods can only apply to the set of FPS and
frequencies seen in the training data, due to the nature of classification algorithms.  We avoid this drawback by employing a unconstrained
based approach -- our regression-based model can be used for arbitrary processor frequencies (even those that were not presented during
training), because the model takes the frequency as an input.

\subsection{Predictive Modeling}

Our model for determining the optimal processor setting is a collection of artificial neural networks (ANNs). \dl{As we target two
interactive events in this work, we have two ANNs - one for each event.} We choose ANNs because they deliver better and more robust
performance than alternative classification techniques like support vector machines and decision trees (see Section~\ref{sec:alter}).

We follow the classical 3-step supervised learning to build and deploy our models: (i) generate training data and problem modeling, (ii)
train a model on training data, (iii) and use the model on unseen data. We describe each of the steps in detail in the following
subsections.

\subsection{Training Data Generation and Problem Modeling}
\begin{figure}[t!]
  \centering
  \includegraphics[width=0.5\textwidth]{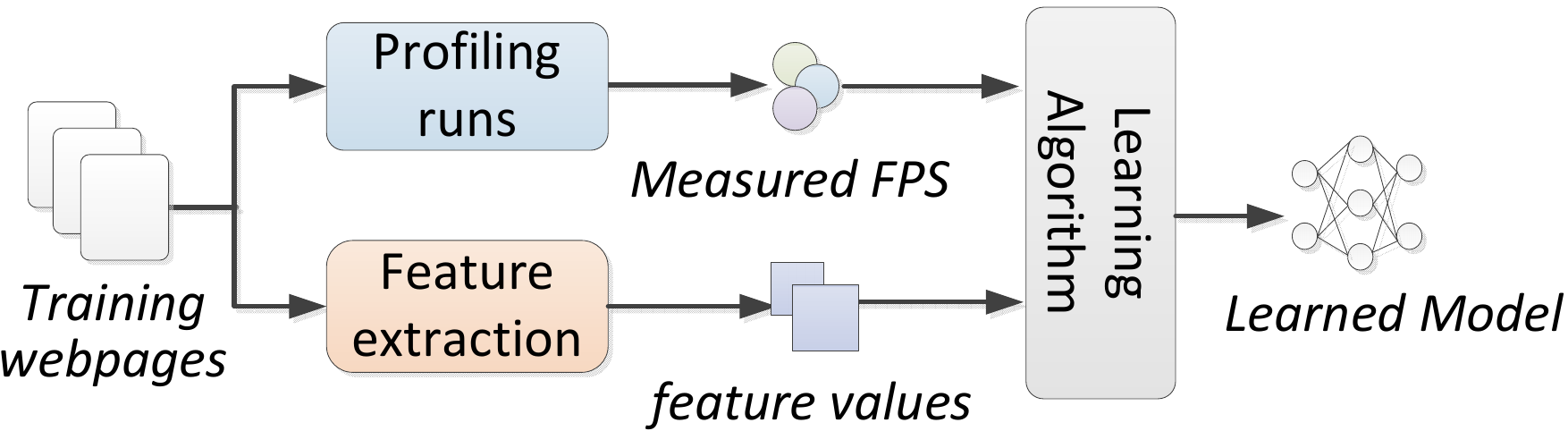}\\
  \caption{Training predictive models. Our models are training off-line so that the user does not pay the profiling cost during runtime deployment.}
  \label{fig:training}
\end{figure}

\subsubsection*{Training data generation}
Figure~\ref{fig:training}  illustrates the process for training an ANN model, which applies to all machine-learning models we evaluate in
this work. Our models are built \emph{offline} using training webpages. In this work, we apply cross-validation by using a corpus of 80
different webpages for training and 20 different webpages for testing (see Section~\ref{sec:method}), \dl{We make sure that the testing webpages are different from
the training webpages and come from a different website.} which are collected from the landing page of the
top-100 hottest websites ranked by \texttt{alexa.com} (see also Section~\ref{sec:web_workload}). \dl{To generate training data, we develop a
script to record and replay the two user events we target.}
For each training webpage, we generate different training scenarios by varying the duration and speed of a target event which is recorded and replayed by a script we developed, we also extract their web feature values from the DOM tree constructed during the page loading phase. In each training
scenario, we exhaustively execute the rendering process under 14 different processor settings and record the achieved FPS. \dl{For each
webpage, we also extract their web feature values from the DOM tree constructed during the page loading phase.}


\hl{More specifically, we automatically generate $8,960$ training samples for each target gesture of one user by varying the process setting,
event rates and initial viewports ($14$ processor settings $\times$ $8$ event rates $\times$ $80$ different training webpages). We found this set of training data to be sufficient for our model
structure.}

\begin{figure}
  \centering
  \includegraphics[width=0.5\textwidth]{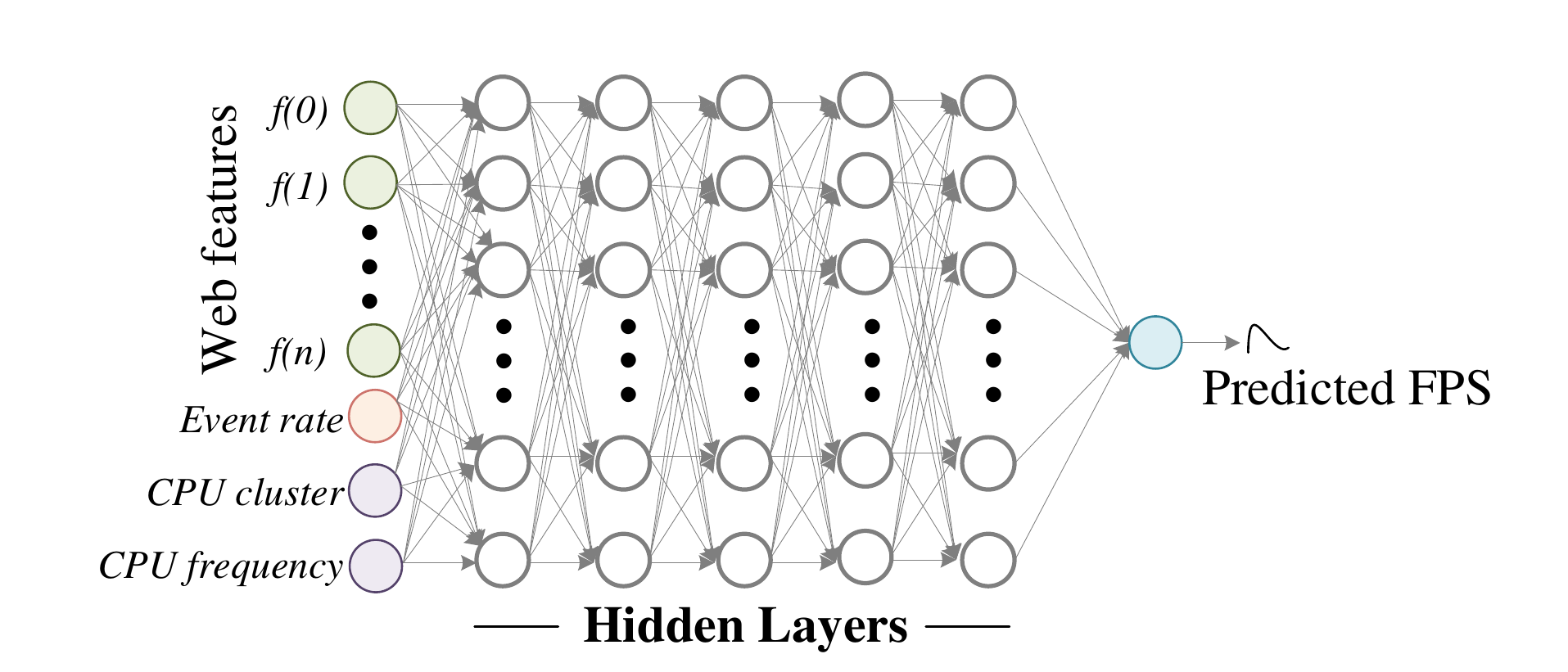}\\
  \caption{Our FPS predictor is a multi-layer neural network.
  The input to the network includes a web-feature vector of real values, the measured incoming event rate,  a label indicates the big or little processor cluster and real value of the processor speed.
The network outputs a real value of the estimated FPS for the given web content with an input event rate under a specific processor setting.
  }\label{fig:modelstruct}
\end{figure}

\subsubsection*{Model structure} Figure~\ref{fig:modelstruct} depicts the architecture of our models, which is a fully connected, feed-forward ANN with
five hidden layers, where each hidden layer has 80 neurons.  \hl{The model structure is determined empirically -- we wrote a script to try
different model structures and choose the best-performing one from our training dataset.} In Section~\ref{sec:alter}, we provide a
quantified analysis on various \dl{neural network} structures.

Our model takes as input the web feature values, a measured event rate, a label indicates where to run the render process (big or little)
and the clock speed of a given CPU cluster. It produces the estimated FPS as a real value. Our output layer is a linear regression function
and we use the rectified linear unit (ReLU) activation function for the hidden layers. \hl{We empirically evaluated three commonly used
activation functions, Sigmoid, Tan and ReLU. The result shows that ReLU gives the best performance in our training dataset. Recent studies in the
machine learning community also suggest that ReLU can better avoid the vanishing gradient problem}~\cite{maas2013rectifier}. We stress that
keeping the network structure simple is essential for achieving fast prediction and for learning an effective model from a relatively small
training dataset.
\begin{table}[t!]
\caption{Raw web features used in the work}
\small
\centering
        \begin{tabular}{rll}
        \toprule
        \multirow{2}{*}{DOM Tree} & \#DOM nodes & depth of tree \\
                & \#each HTML tag & \#each HTML attr. \\
        \rowcolor[gray]{.92}  & \#rules  & \#each property \\
        \rowcolor[gray]{.92}  \multirow{-2}{*}{Style Rules} & \#each selector pattern & \\
        Other  & size of the webpage (Kilobytes) & \\
        \bottomrule
        \end{tabular}
\label{tab:feature}
\vspace{-4mm}
\end{table}

\subsubsection*{Model Features}
\label{sec:features} One of the key aspects in building a successful predictor is finding the right features to characterize the program
space and the input. \hl{Our model takes in three sets of inputs, a vector of web feature values, a numerical value for the input event
rate, a label of the CPU cluster (big or little) and the clock frequency of the CPU cluster (a real value). Web features are used to
capture the workload characteristics, and the event rate and CPU frequency settings directly affect the resultant FPS. }

\hl{To determine what web features are important for characterizing the input web content}, we started from 1,084 raw features that can be
collected at runtime from Google Chromium. Table~\ref{tab:feature} groups our raw web features into categories. The features were chosen
based on previous work on optimizing mobile web browsing~\cite{Ren:2018:PNW:3281411.3281422}, as well as our intuitions.

\cparagraph{Feature reduction} To learn a useful model, supervised learning typically requires the number of training samples to be an
order of magnitude larger than the number of model inputs (i.e., features). Given that our training dataset size (i.e., 80 webpages) is
less than the number of raw features, we need to find ways to reduce the dimensionality of the feature space. We do so by first applying
Principal Component Analysis (\PCA)~\cite{dunteman1989principal}  to the raw features, and then choosing the top 49 principal components
(\PCs) which account for around 95\% of the variance of the original feature space. We record the \PCA transformation matrix and use it to
transform the raw features of the new webpage to \PCs during runtime deployment. \PCA is a standard statistical method for reducing the
dimensionality of data. By reducing the feature dimension, we are also improving the generalizability of our models, i.e. reducing the
likelihood of over-fitting on our training data.

\cparagraph{Feature normalization} Before passing the extracted feature values to a machine learning model, we normalize or scale each of
the features to a common range (between 0 and 1)  to prevent the range of the order of the feature value is a factor in its importance.
Scaling features does not affect the distribution or variance of their values. To scale the features of a new webpage during deployment, we
record the minimum and maximum values of each feature in the training dataset and use these to scale the corresponding features. If an
extracted value of an unseen webpage is outside the min-max range, we clip it to the range.

\begin{figure}[t!]
  \centering
  \includegraphics[width=0.4\textwidth]{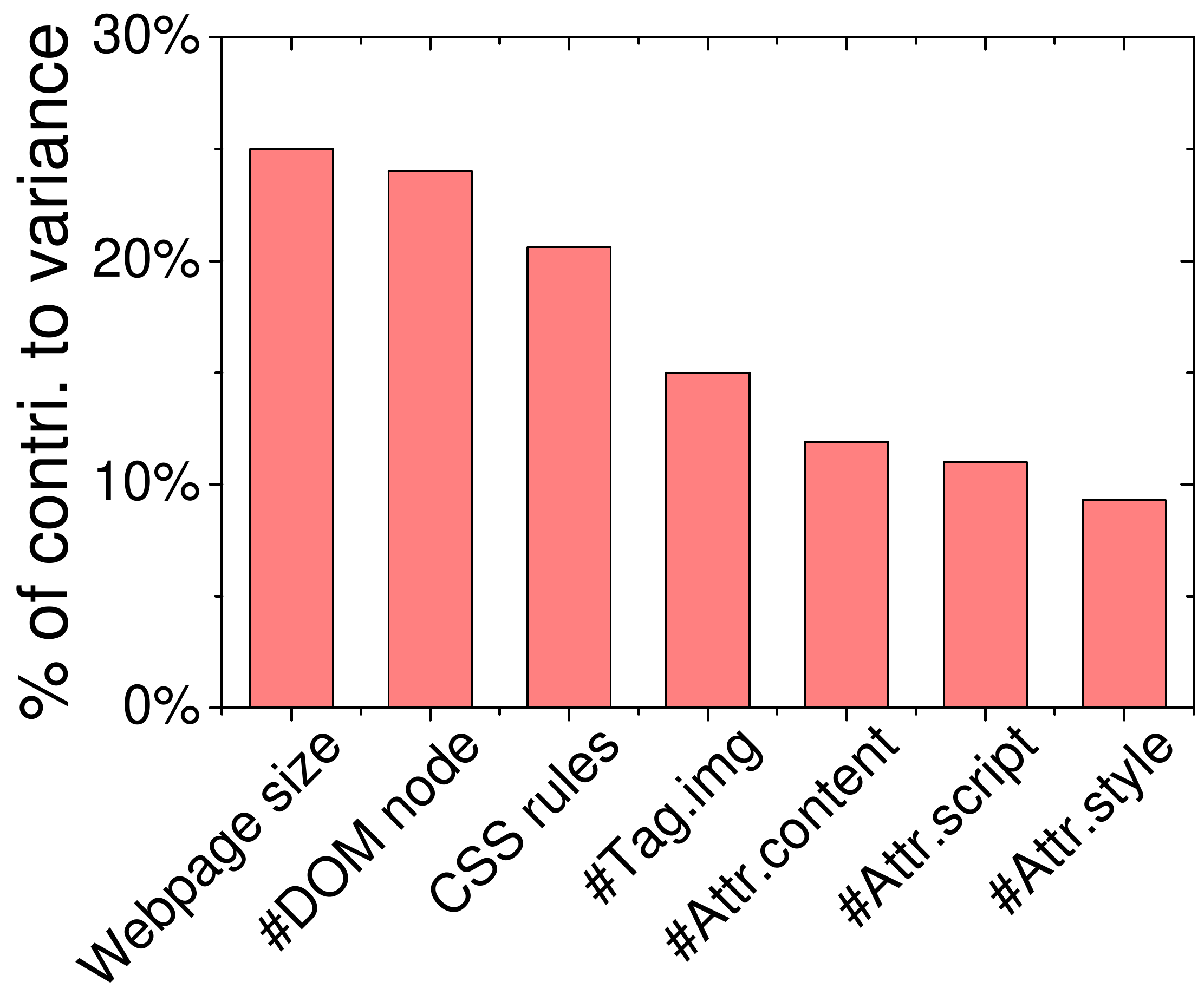}\\
  \caption{Top 7 most important raw web features after applying Varimax
rotation and their contributions to the variances in the PCA space.}
  \label{fig:pca_importance}
\end{figure}

\cparagraph{Contributions of raw features} To obtain some insights for the usefulness of each raw feature, we apply the Varimax
rotation~\cite{manly2016multivariate} to the feature space after applying PCA. This technique quantifies the contribution of each feature
to each PC in terms of variances. Figure~\ref{fig:pca_importance} shows the top 7 dominant features based on their contributions to the
PCs. Features like the webpage size and the number of DOM nodes make significant contributions to the PCA space and are hence considered to
be important. This is not surprising because the larger the webpage size and the number of DOM nodes are, the more processing time will be.
Other features, like \# CSS rules, and \# Tag.img, also make great contributions to the variance on the PCA space. This because they
determine how the webpage should be presented and how do they correlate to the rendering overhead. By employing an automatic feature
selection and tuning process, our approach has the advantage of having better portability when targeting a new hardware architecture where
the cost of web processing and the importance of web features may change. Later in Section~\ref{sec:importance}, we provide a further
analysis on the feature importance via a Hinton diagram.


\subsection{Model Training}

The feature values of the target web content, the event speed, and the processor frequency together with the measured FPS are passed to a
supervised learning algorithm to learn an ANN for each targeting event. The learning algorithm then tries to update the weights of the ANN
to closely map the model input to the measured FPS.

Our models are trained using back-propagation with stochastic gradient descent (SGD). For a set of training examples $X_1 \ldots X_n$, the
SGD algorithm tries to find a set of network parameters $\Theta$ that minimize the output of a loss function:

$$
\Theta = \argmin_{\Theta} \frac{1}{n} \sum_{i=1}^{n} \ell \left(X_i, \Theta \right)
$$

where loss function $\ell \left(x, \Theta \right)$ computes the mean squared logarithmic error between the model's outputs, $\hat{x}$, and
expected values, $x_i$:

$$
\ell = \frac{1}{n} \sum_{i=1}^{n} (log(x_i + 1) - log(\hat{x_i} + 1))^2
$$

We choose this loss function because it penalizes underestimates more than overestimates. This reduces the chance of QoS violations due to
an underestimated FPS target. Our model is trained using the Adam learning algorithm~\cite{kingma2014adam}.

In this work, we train an ANN for each type of events. Since we target two types of events, scrolling and pinching, we build two ANNs. It
is to note that an alternative is to have a single model for all event types. However, this strategy offers little flexibility for updating
and extension as doing so would require retraining the whole model when targeting a new event. Furthermore, this alternative strategy not
only will incur expensive re-training overhead but also is likely to be less effective than a specialized model~\cite{marco2017improving}.


\begin{figure*}[t!]
  \centering
  \includegraphics[width=1\textwidth]{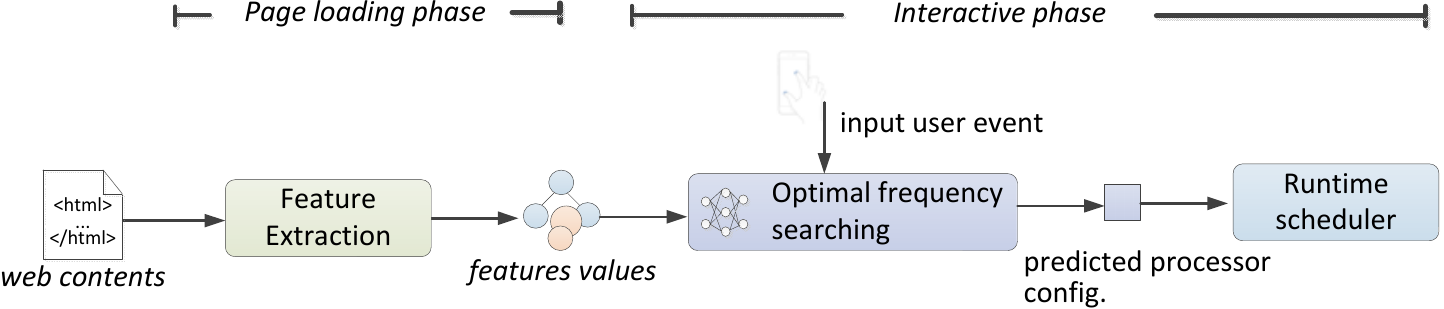}\\
   \vspace{-2mm}
  \caption{Using the trained predictive model to find the optimal processor configuration during the interactive phase.}\label{fig:deployment}
   \vspace{-2mm}
\end{figure*}













\begin{algorithm}[t!]
\caption{Processor setting search engine} \label{alg:search}
\SetAlgoLined \KwResult{$c_{opt}$: desired processor setting}
\KwIn{$FPS_{min}$: minimum acceptable FPS;}
\KwIn{$r$: event rate;}
\KwIn{$C[0..N-1]$: available processor settings, sorted by frequencies from low to high;}

low = 0;

high = N-1;

\While{low <= high}{
    mid = (low + high)/2;

    $FPS_{pred}$ = $pred(C[mid], ...)$;

    \uIf{$FPS_{pred}$ $>$ $FPS_{min}$}
       { high = mid - 1; }

    \uElseIf{$FPS_{pred}$ $<$ $FPS_{min}$}
       { low = mid + 1;}

    \Else
       { \Return $c_{opt} = C[mid]$;}
  }
       $FPS_{low} = pred(C[low], ...)$;

     $FPS_{high} = pred(C[high], ...)$;

     /*Return the closest FPS setting*/

     \Return $c_{opt} = (FPS_{low} - FPS_{min}) < (FPS_{min} - FPS_{high}) ? C[low+1] : C[high]$;
\end{algorithm}

\subsection{Model Deployment}

Our models are implemented in the Python scikit-learn machine learning package. The trained models are encapsulated in a Python library to
be invoked by the web browser (via a browser extension in our prototype) for any webpage that is not seen in the training phase. For this
work, we have developed a working prototype based on the Chromium. Our implementation requires small changes to the web browser - in total,
we have modified around 700 lines of code.

Figure~\ref{fig:deployment} shows how the trained models can be used during the interactive phase to determine the processor clock
frequency. Feature values are extracted from the DOM tree, during the page loading phase after the downloaded web contents are parsed to
construct the DOM tree. The extracted feature values are re-used throughout the interactive stage unless the DOM tree has changed
significantly due to e.g., content reloading. Specifically, if there is more than 30\% difference in the number of nodes between the
previous and the current DOM trees, we will update the feature values by performing feature extraction on the current DOM tree. The
prediction and frequency configuration will be triggered if one of the targeting user input is detected. To make a prediction, we first
choose a model for the input event. The chosen model is then used to estimate the achieved FPS under different processor settings to find
out the optimal setting. The predicted setting is passed to the runtime scheduler to perform task scheduling and hardware configuration. We
note that the runtime scheduler only reconfigures the hardware if the predicted setting is different from the current one.

\hl{Our model also takes as input the event rate. This is calculated based on a sampling window of 200 ms. This window is also used to
smooth the input event speed. We found that all the interactive sessions captured in our user study fit into this window. Furthermore, we
do not predict future event rates, but a change of the event rate in the next sampling window might trigger a reconfiguration of the
processor settings.

We also note that a frequently changed event rate might affect the response time and energy savings. However, we found that this rarely
happens (during an interactive session) for the two events we target (scrolling and pinching). }

The pseudocode in Algorithm~\ref{alg:search} describes our binary-search-based processor setting search algorithm. The search engine uses
the predictive model (line 5), $pred$, to quickly find a desired processor setting, $c_{opt}$,  from a range of available options, $C[]$. The
goal is to find a processor configuration which hopefully will lead to an FPS that is as close as possible to the minium acceptable FPS,
$FPS_{min}$. It is possible that none of the predicted FPS values, $FPS_{pred}$, exactly matches the minimum acceptable FPS, $FPS_{min}$.
In this case, we return the one that gives the closest FPS value (line 17), but we always choose the next higher frequency setting,
$C[low+1]$, to increase the likelihood for meeting $FPS_{min}$. \hl{In the extreme case, we might choose to use the highest clock frequency
provided by the hardware.}

It is also worth mentioning that the overhead of feature extraction, model prediction, and processor frequency searching and configuration
is small. It is less than 10 ms which is already included in our experimental results.

\section{Evaluation Setup}
\label{sec:setup} We now describe our experimental setup and evaluation methodology.

\begin{table*}[t!]
\begin{center}
\caption{Mobile platforms used in evaluation}
\label{tbl:hardware}
\begin{tabular}{lll}
\toprule
&\textbf{Odroid Xu3}& \textbf{Nvidia Jetson TX2}\\
\midrule
\rowcolor{Gray} \textbf{big CPU} & 32bit quad-core Cortex-A15 @ 2.0GHz  &  64bit quad-core Cortex-A57 @ 2.0 GHz\\
\textbf{LITTLE CPU} & 32bit quad-core Cortex-A7 @ 1.4GHz &  64bit dual-core Denver2 @ 2.0 GHz \\
\rowcolor{Gray}  \textbf{GPU} &  8-core Mali-T628 @ 600MHz  & 256-core NVIDIA Pascal @ 1.3GHz \\
\bottomrule
\end{tabular}
\end{center}
\end{table*}

\subsection{Hardware and Software Platforms}
We evaluate our approach on two distinct mobile platforms: an Odroid Xu3 and a Jetson TX2. Table~\ref{tbl:hardware} gives detailed
information about the evaluation platforms. Both platforms implement the widely used ARM big.LITTLE mobile architecture but with different
CPU generations and frequency setting knobs. The Odroid Xu3 implements the  Exynos 5410 SoC that was released in 2014, and thus represents
a low to medium end mobile spec. It is to note a recent study published in 2019~\cite{wu2019machine} suggests that 75\% of today's
smartphones still use a CPU design that was released before 2013. Therefore, including Odroid Xu3 in our evaluation ensures that our
approach is evaluated on a platform that presents a wide range of mobile devices. In contrast to Odroid Xu3, the Jetson TX2 integrates a
more recent SoC (released in 2017), and has larger RAM and more powerful CPUs. Therefore, it represents a higher end, more recent
smartphone spec.

We use the onboard energy sensors provided by both systems to measure the power consumption of the \emph{entire} system. These sensors and
power meters have been proven to be accurate in prior work~\cite{Ren:2018:PNW:3281411.3281422}.

For systems software, our evaluation systems run Ubuntu 16.04 with the big.LITTLE enabled scheduler\footnote{Because Chromium for Android
does not support extensions, we implemented our approach on the Linux version that shares the same code base as the Android Chromium and
\hl{compare to the Android default interactive CPU frequency governor}. However, our techniques can be built directly into the Chromium
browser for Android, and we leave this as our future work.}. We implemented our approach in Google Chromium (ver. 73.0) which is compiled
using the gcc compiler (ver. 7.2) with default compilation options provided by the Chromium using the ``release" version.


\begin{figure}[!t]
	\centering
	\subfloat[][]{\includegraphics[width=0.25\textwidth]{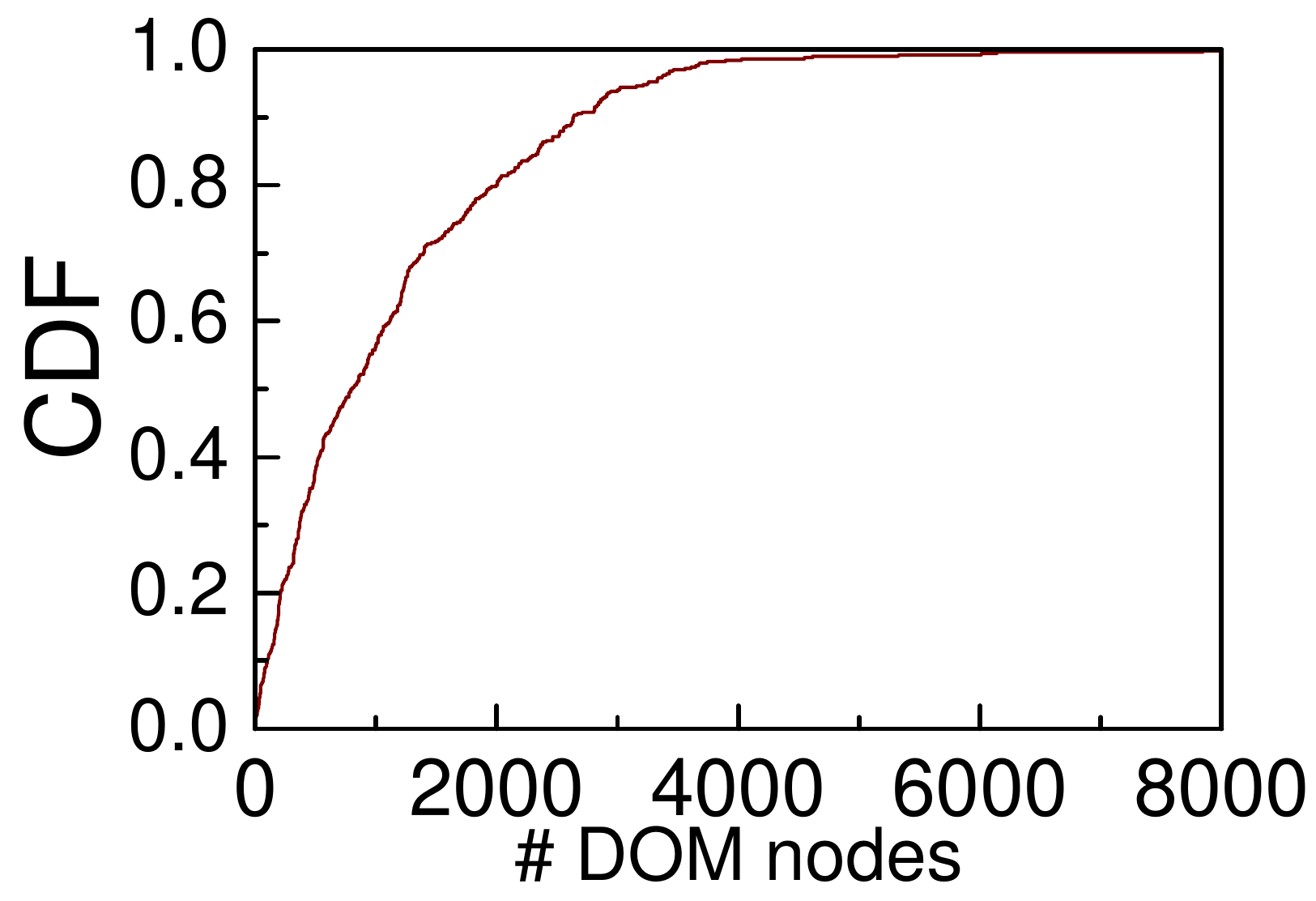}}
    \hspace{-5mm}
    \hfill
        \hspace{-5mm}
    \subfloat[][]{\includegraphics[width=0.25\textwidth]{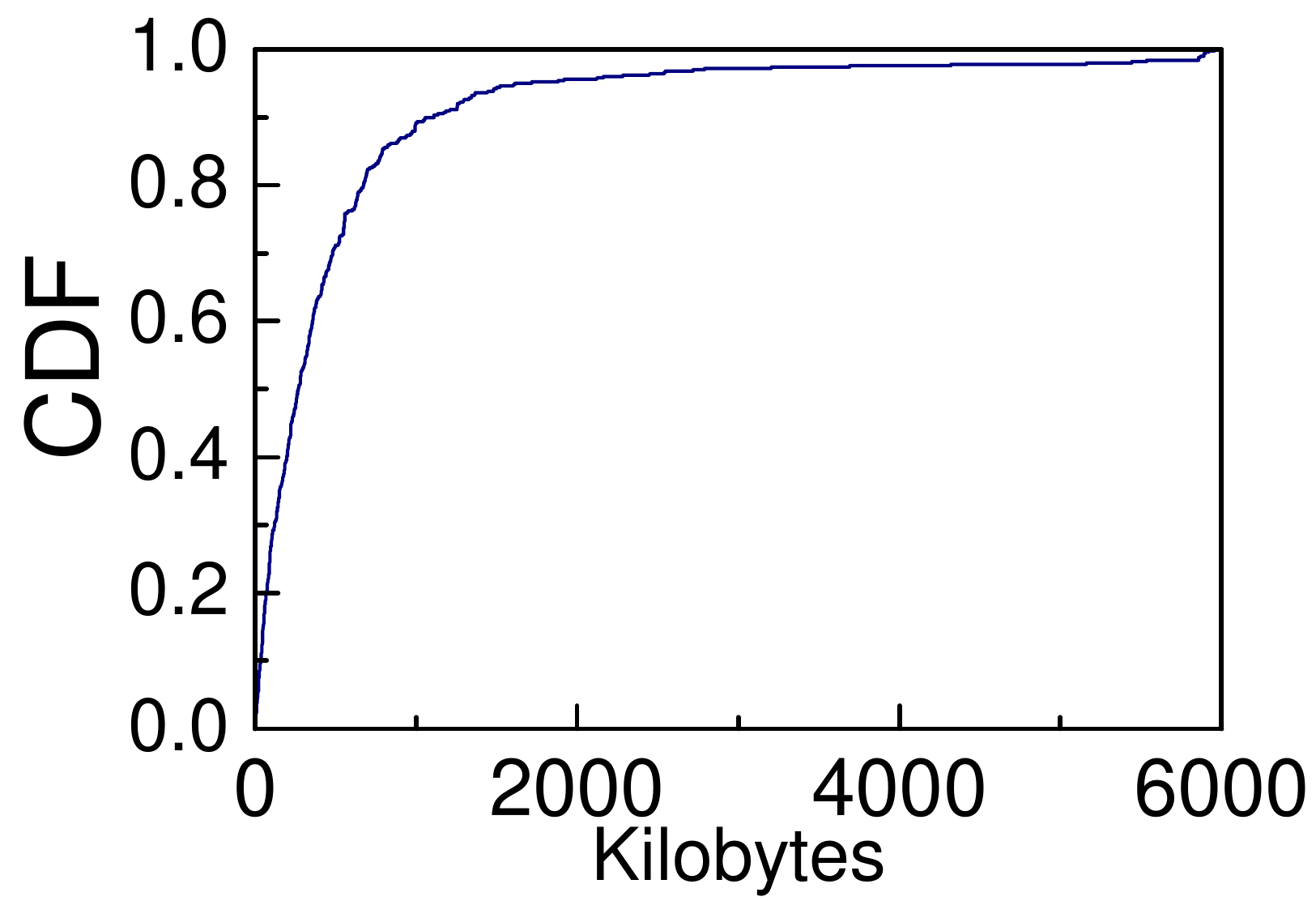}}
    \caption{The cumulative distribution function (CDF) for the number of DOM nodes (a) and size of web content (b) of the 100 webpages used in evaluation. }
    \vspace{-3mm}
    \label{fig:webfeature}
\end{figure}
\subsection{Web Workloads}
\label{sec:web_workload} Throughout this work, we use the landing page of the top 100 hottest websites (as of April 2019) from
\texttt{www.alexa.com}. We use the mobile version of a website if available. Figure~\ref{fig:webfeature} shows the CDF of the number of DOM
nodes and web content sizes. The DOM node and webpage sizes range from small (4 DOM nodes and 40 KB) to large (over 8,000 DOM nodes and 6
MB). The wide distribution of webpages indicates that our test data cover a diverse set of webpages. \hl{Overall, we test our approach on
16,000 samples ($100$ webpages $\times$ $20$ users $\times$ $8$ event rates), representing one of the largest-scale experiments seen to
date on mobile web browser optimizations.}

\subsection{Baseline and Competitive Approach}
\cparagraph{Baseline} As a baseline, we use \texttt{interactive} as the default CPU frequency governor. This is a standard power management
policy used by the Android system for interactive applications. We use the default setting of the interactive governor, described as
follows. The governor samples the CPU load within a window of 80 ms. It raises the frequency if the CPU utilization is above 85\%; after
that, it waits for at least 20 ms before re-sampling the CPU to decide whether to lower or raise the frequency.

\cparagraph{State-of-the-art} We compare our approach against eBrowser~\cite{xu2018ebrowser}, the most closely related recent work.
eBrowser reduces the energy consumption for a given user event by putting the rendering process into sleep for some time. This essentially
reduces the number of events to be processed as some of the user events within an interaction window will be dropped by the browser during
sleep. eBrowser uses a linear regression model to model the acceptable event rate on a per-user basis. However, it requires statistical
data to be sent to a remote server to learn a model and relies on the operating system for power management. By contrast, our approach does
not drop user events (as doing so could miss important inputs) and actively participates in power management by using the knowledge of the
web workloads to determine the processor configuration. \hl{We control the FPS by adjusting the processor running frequency. For example, a
higher CPU frequency will enable the system to process more frames per second.} We port the open source implementation of
eBrowser\footnote{\url{ https://github.com/cloud-ecnu/ebrowser-1}} to the latest version of Chromium used in our experiments.

\subsection{Evaluation Methodology}
\label{sec:method} \cparagraph{Predictive model evaluation}  \hl{We use five-fold cross-validation in our experiments. Specifically, we
partition our 100 webpages into 5 sets where each set contains 20 webpages. We keep one set as the validation data for testing our model,
and the remaining 4 sets for training data to learn a model. We repeat this process five times (folds) to make sure that each of the 5 sets
used exactly once as the validation data. We then report the averaged accuracy achieved across the 10 validation sets}. This is a standard
evaluation methodology, providing an estimate of the generalization ability of a machine-learning model in predicting \emph{unseen} data.

\cparagraph{Evaluation metrics}In our evaluation, we use two metrics: energy reduction and QoS violation. Energy reduction is normalized to
the energy measurement when using the \texttt{interactive} CPU governor. QoS violation is calculate as $\delta/FPS_{min}$, where $\delta$
is the number of FPS falls below the minimum acceptable FPS, $FPS_{min}$. If the resulting FPS is greater than the minimum acceptable FPS,
we consider there is no QoS violation (but this may lead to higher energy consumption when reporting energy saving).

\cparagraph{Measurements} To measure energy consumption, we developed a lightweight runtime to take readings from the onboard energy
sensors at a frequency of 100 samples per second. We then matched the energy readings against the timestamps in an interactive window to
calculate the energy consumption. For the FPS, we develop a web extension to record the number of request calls processed by the browser
per second, by counting the number of invocations of the Chromium \texttt{ window.requestAnimationFrame()} API.

\cparagraph{Performance report} Unless state otherwise, we report the \emph{geometric mean} across experimental settings. We note that
geometric mean has been shown to be better at minimizing the impact of performance outliers over arithmetic mean, and is a preferred metric
for performance reporting~\cite{ertel1994definition}. To collect run-time and energy consumption, we run each approach on a testing input
repeatedly until the variance under a 95\% confidence per input is smaller than 2\%. This repeat running strategy is essential for
obtaining statistically sounded results. Finally, to isolate the impact of network latency, all the testing webpages are downloaded and
loaded from the disk. We also disable the cache of the web browser to ensure consistent results across different runs of the same page. We
consider this is a reasonable setting as our work focuses on the interactive phase where most of the content would have already been
downloaded.

\subsection{Quantifying QoS}
\label{sec:qqos}
\begin{figure}[t!]
  \centering
  \includegraphics[width=0.45\textwidth]{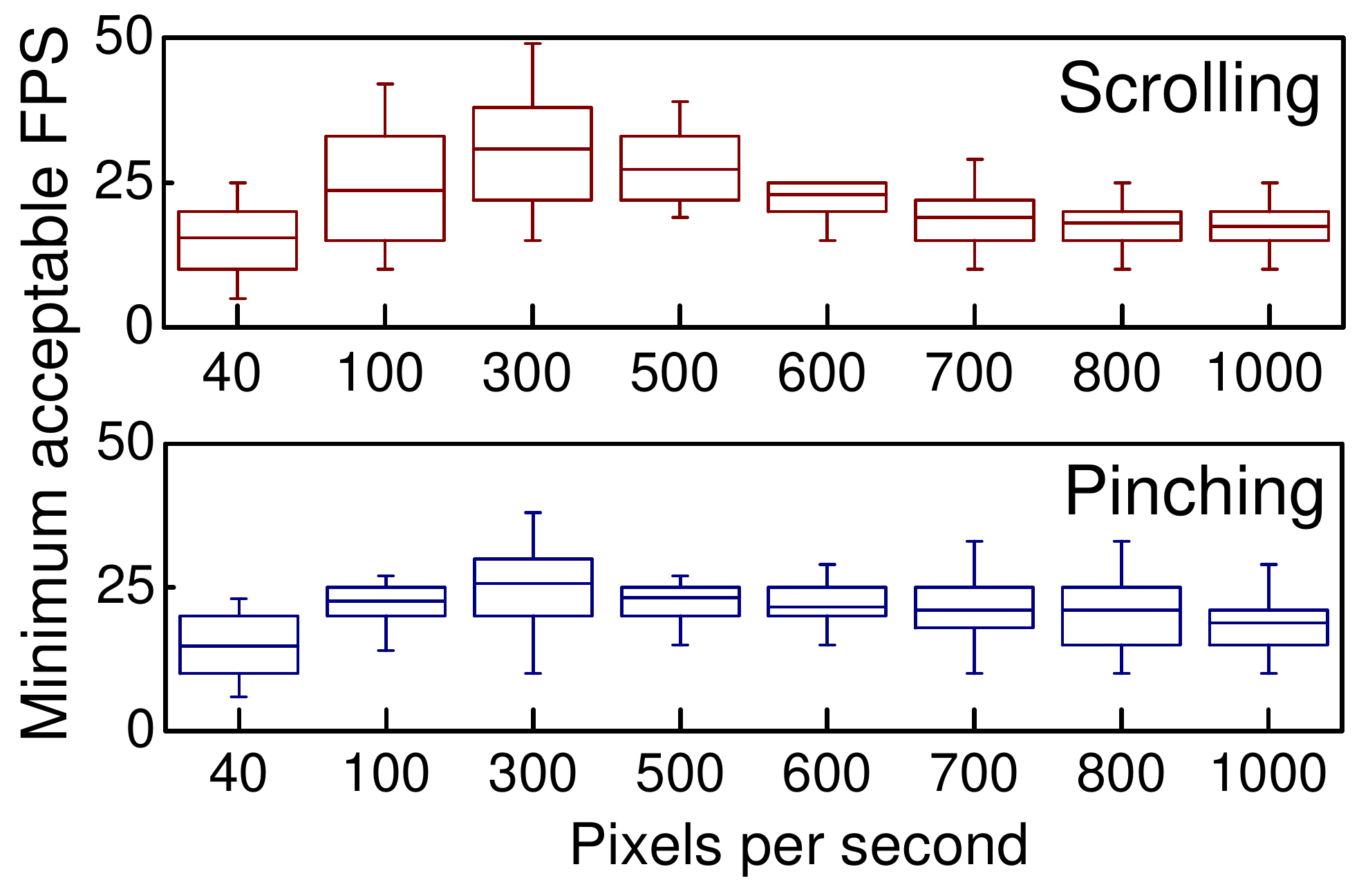}\\
  \caption{The minimum acceptable FPS across 100 webpages, 20 users and eight event rates (measured by the number of pixels per second touched by the
  finger). The ``whisker" of a box shows the variance across webpages and users. The minimum acceptable FPS changes across event
  rates, webpages and users, suggesting an adaptive optimization scheme is needed. }\label{fig:minFPS}
\end{figure}

 To quantify the QoS during web interactions, we conducted a user study. Our user study involved 20 participants (10
females) who were the students at our institution during the time this work was conducted. The participants were at the age group of under
30 and are a frequent user of web-related mobile applications. In our experiment, we automatically replay the user interactions on 100
webpages for each of the targeting gestures (scrolling and pinching) under eight event rates (quantified by the number of pixels per second
touched by the  finger). In this user study, we display the content under various on-screen update speed (measured by the FPS).  We then
ask each user to score the experience using a Likert Scale of 5 scores, where a score of 0, 3 and 5 being very dissatisfied, acceptable and
very satisfied respectively.

Figure~\ref{fig:minFPS} plots the minimum acceptable FPS for scrolling and pinching, averaging across testing webpages and users. The
mini-max bar shows the variation across different users. The x-axis shows the eight event rates used in the evaluation. \hl{We stress that
our approach can be applied to an arbitrary event rate as the measured event rate is part of the model's input.}
 Our user study suggests that the FPS strongly correlates to
the QoS. For the same user, the acceptable QoS for a given event-rate for a gesture corresponds to more or less the same FPS (with a
standard deviation of less than 4.4). However, the minimum acceptable FPS varies across users and events, indicating adaptive optimization
is required. \hl{Note that} our observations are in line with the findings reported by \hl{eBrowser}~\cite{xu2018ebrowser} and \hl{we do
not claim novelty in this aspect}.

In our experiments, we use the results of this user study as the minimum acceptable FPS guidelines. When reporting QoS violations, we
measure the performance of each scheme for each testing page under each user-specific acceptable FPS. For reproducibility, in our
experiments, we automatically generate eight different event rates for each testing page using a script. We then report the performance of
energy saving and QoS violations across 100 webpages, 20 user-specific minimum acceptable FPS settings and eight event rates.

\section{Experimental Results}
In this section, we first report the overall results of our experiments, showing that our approach consistently outperforms the
state-of-the-art across hardware architectures and evaluation metrics. We then provide details on the working mechanism of predictive
modeling, including the prediction accuracy and distribution, feature importance, overhead and alternative modeling techniques.

\begin{figure*}[t!]
  \centering
  \vspace{-2mm}
 \subfloat[][Average energy reduction]{
  \includegraphics[width=0.26\textwidth]{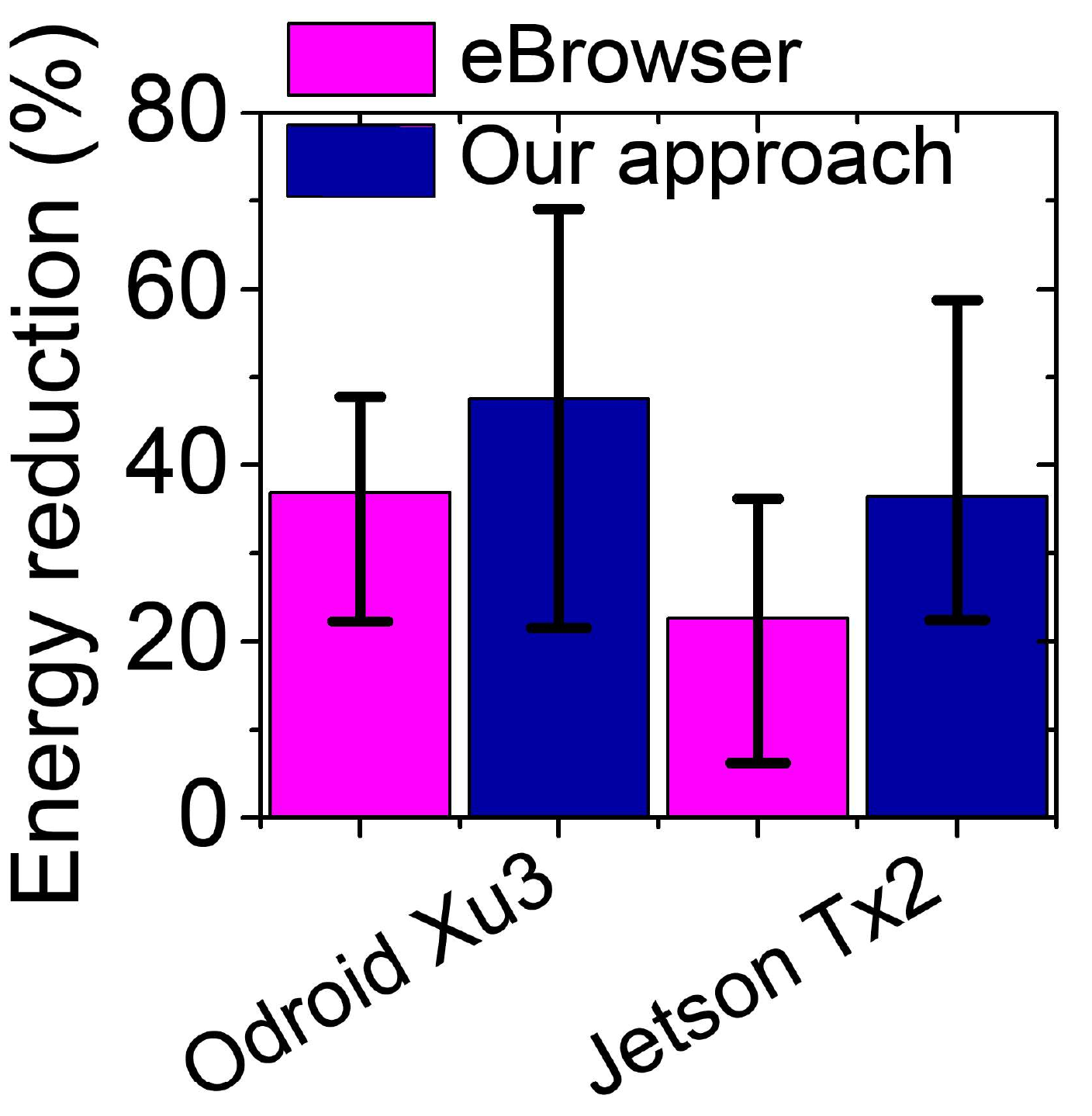}
  }
 \subfloat[][Distribution of energy reduction]{
    \includegraphics[width=0.36\textwidth]{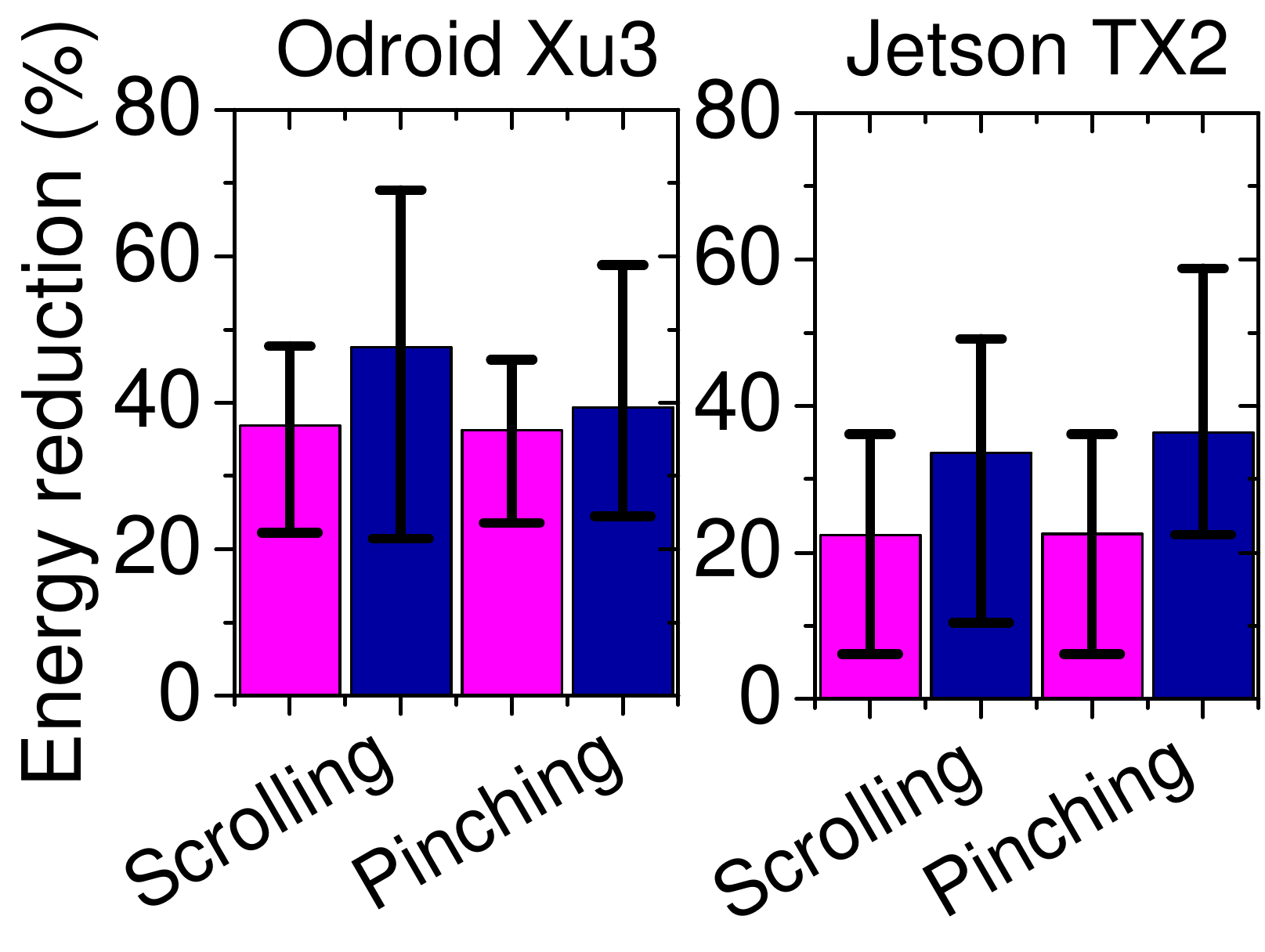}
 }
 \subfloat[][Distribution of QoS violations]{
    \includegraphics[width=0.37\textwidth]{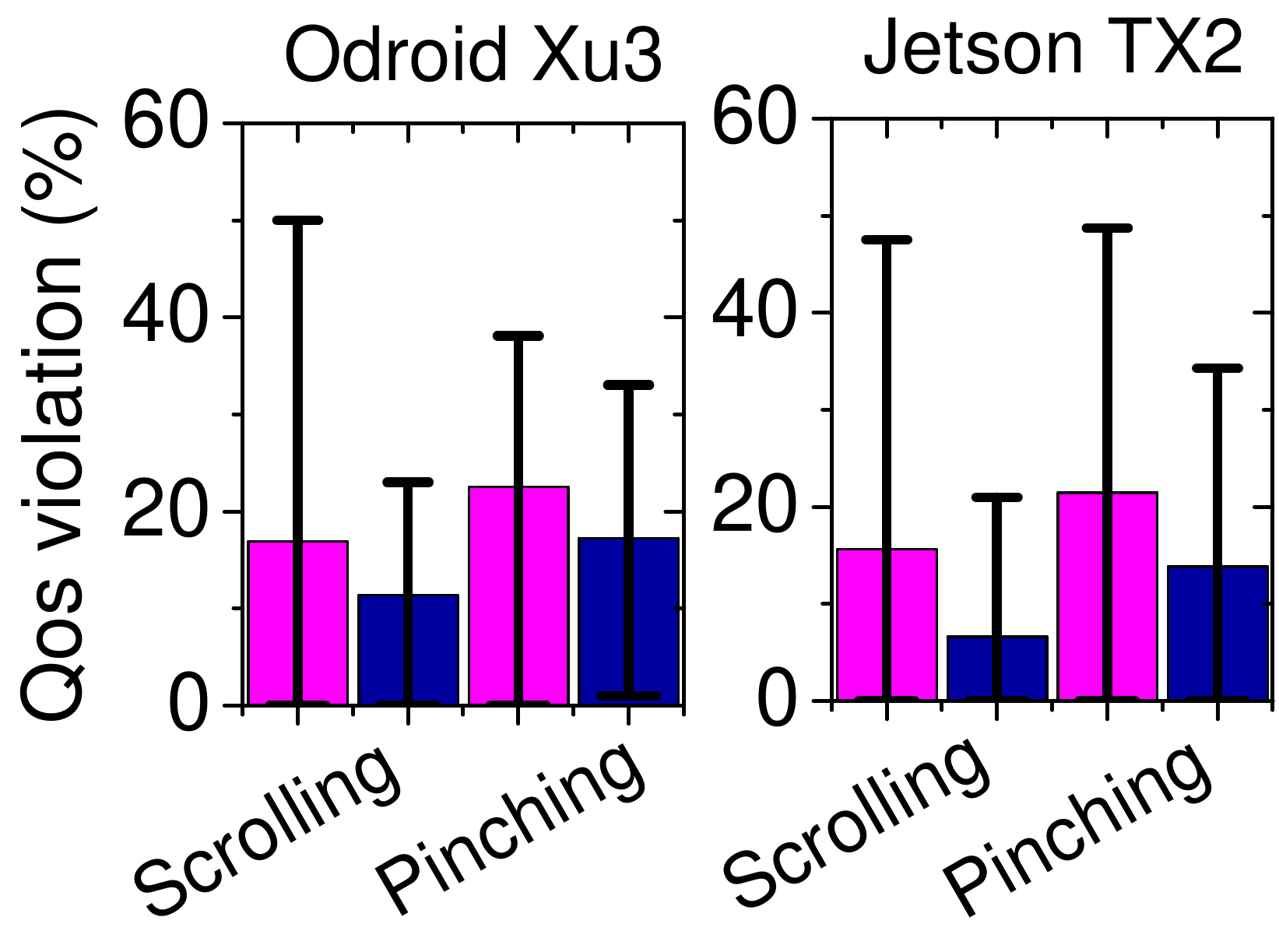}
 }

  \caption{Average (a) and the distribution of energy reduction (b), as well as the distribution of QoS violations (lower is better) over the
  \texttt{interactive} CPU governor across 100 webpages for meeting the QoS targets across 20 users. Our approach achieves over 12\% more
  energy saving but with less frequent QoS violations.}\label{fig:energyReduction}
\end{figure*}

As a highlight, our key findings are:

\begin{itemize}
\item Our approach delivers consistently more energy saving but with a lower QoS violation when comparing to the state-of-the-art on both
    of our evaluation platforms (Section~\ref {sec:overallr}).
\item Our approach gives consistent good performance for predicting the resultant FPS under a given processor setting, with a low average
    prediction error of less than 15\% (Section~\ref {sec:predaccuracy}).
\item We provide a detailed analysis of the working mechanism of our approach to justify the design choices
    (Sections~\ref{sec:importance} to \ref{sec:alter}).

\end{itemize}

\subsection{Overall Performance}
\label{sec:overallr} Figure~\ref{fig:energyReduction}a compares the energy reduction of our approach against eBrowser on Odroid Xu3 and
Jetson TX2, where the baseline is the default \texttt{interactive} CPU frequency governor. The min-max bars show the variance of energy
reduction. By trading responsiveness for energy, both approaches were able to lower the energy consumption for processing user events.
eBrowser gives an average energy reduction of 36.9\% and 22.6\% on Odroid Xu3 and Jetson TX2 respectively. By exploiting the processor
frequency and heterogeneous architecture design, our approach gives a higher energy saving of 47.6\% (up to 70\%) and 36.4\% (up to 60\%) on
Odroid Xu3 and Jetson TX2 respectively. These translate to an improvement of 17\% and 17.8\% on energy reduction over eBrowser on Odroid
Xu3 and Jetson TX2 respectively.

Figure~\ref{fig:energyReduction}b shows the distribution of energy reduction across testing webpages for each of our evaluation platforms.
The min and max bars represent the highest and the lowest energy reduction found across 100 webpages for meeting the QoS metric of 20
users. Our approach consistently outperforms eBrowser not only with a larger averaged energy reduction but also with a better improvement
for 80\% of the webpages. On only 20\% of the webpages, our approach gives marginally lower energy savings (less than 10\%), but our
approach does not miss or drop any user event like eBrowser.

Figure~\ref{fig:energyReduction}c compares the QoS violation of our approach against eBrowser on both evaluation platforms for scrolling
and pinching. We observe a higher QoS violation for pinching over scrolling. This is because scrolling often lasts longer than pinching, which
offers more room for scheduling and predictions. While eBrowser can reduce the energy consumption by processing fewer user inputs, it
incurs an average QoS violation of 19.5\% (up to 52.3\%) and 19\% (up to 47.5\%) on Odroid Xu3 and Jetson TX2 respectively. By contrast,
our approach has a lower QoS violation of  less than 12.5\% and 16\% on Odroid Xu3 and Jetson Tx2 respectively. This suggests that our
approach can reduce energy consumption while can maintain a higher level of QoS compared to eBrowser.

\begin{figure}
  \centering
  \subfloat[][Odroid Xu3]{\includegraphics[width=0.5\textwidth]{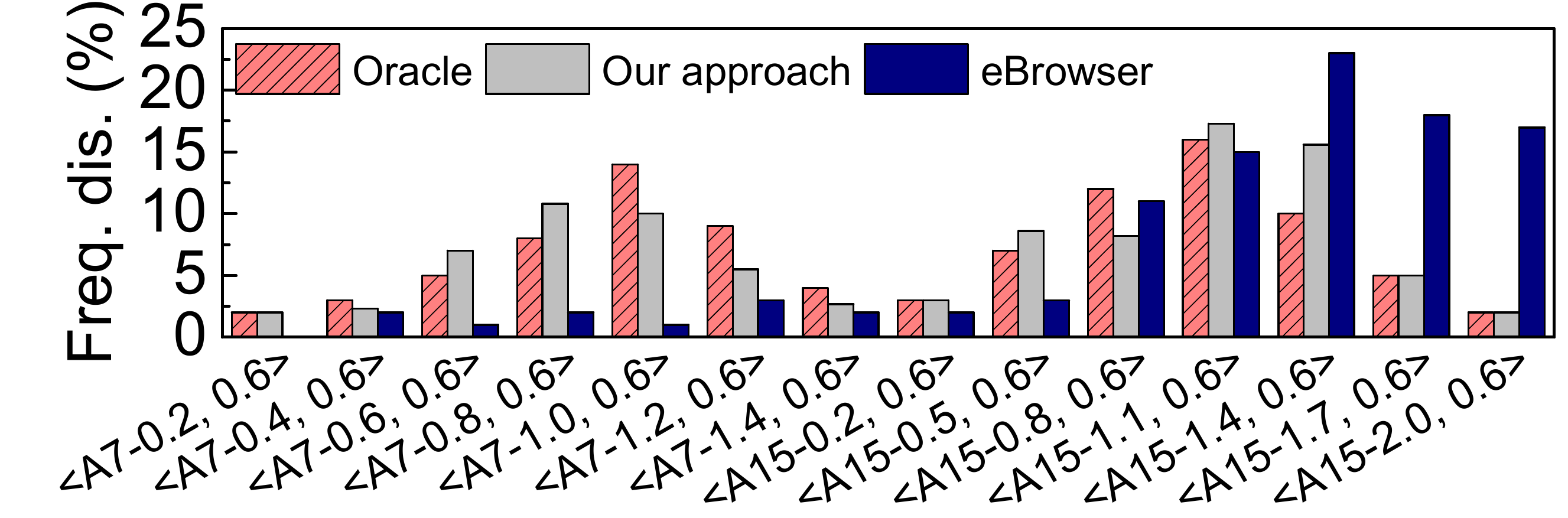}}\\
  \subfloat[][Jetson TX2]{\includegraphics[width=0.5\textwidth]{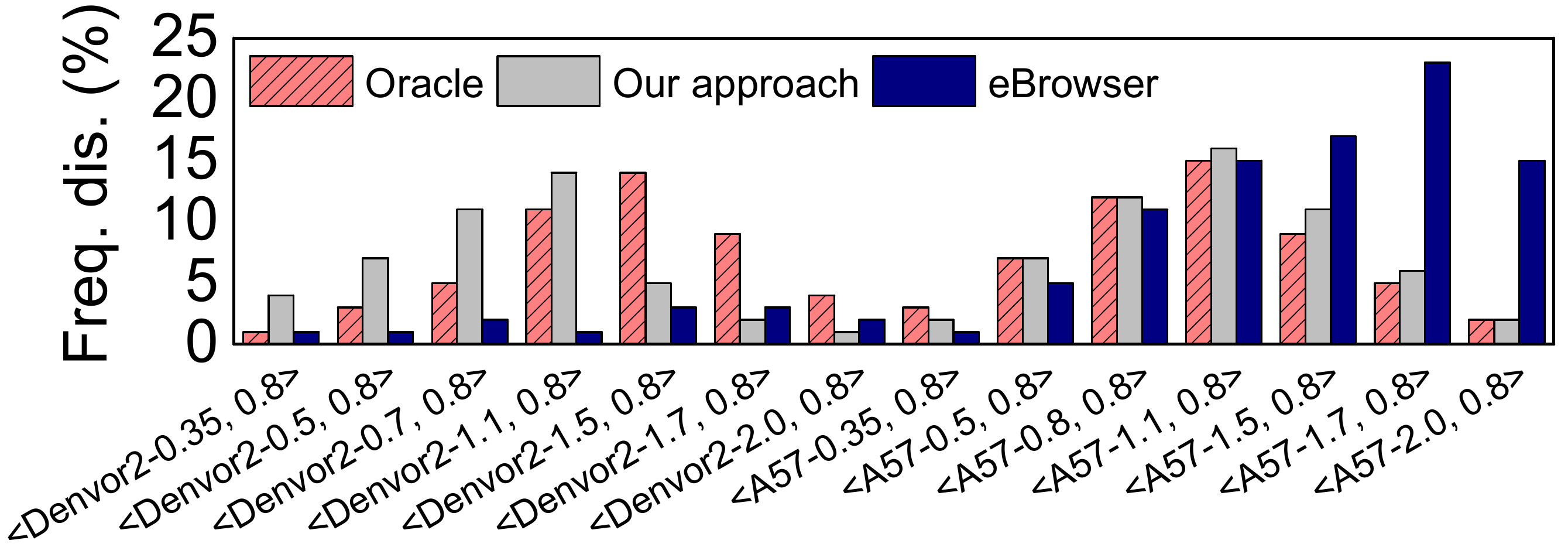}}
  \caption{\hl{The distribution of processor configuration chosen by an oracle approach, our approach and eBrowser. Our approach chooses the low-power processor frequencies more often than eBrowser, which
  gives higher energy savings.}
  }\label{fig:conf_dis}
\end{figure}

\begin{figure}[t!]
  \centering
   \vspace{-2mm}
  \includegraphics[width=0.35\textwidth]{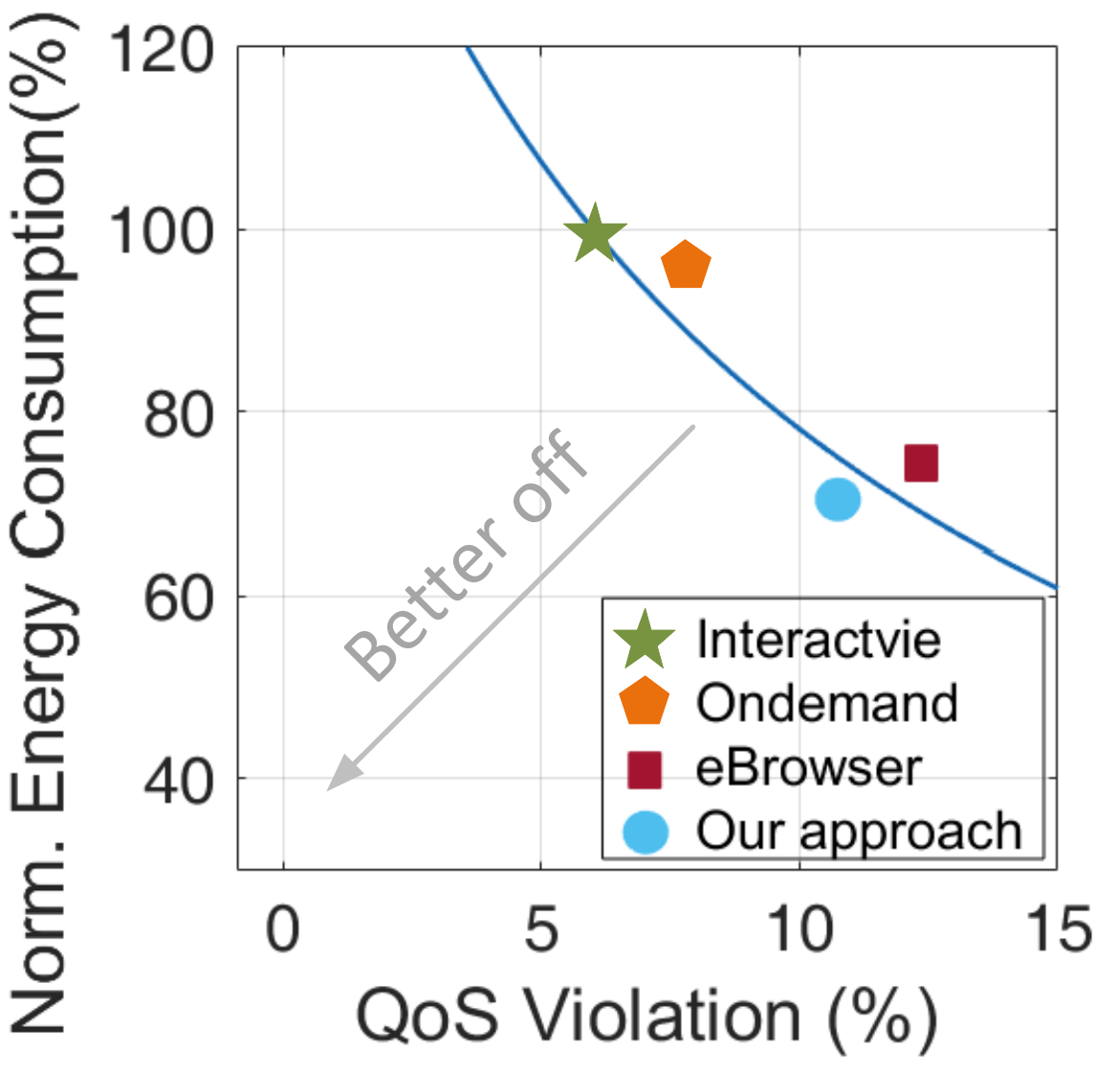}\\
  \caption{A Pareto efficiency diagram shows the QoS-energy tradeoff given by different scheduling policies when interacting with the
  landing page of \texttt{cnn.com} (Section~\ref {sec:motivation}). Energy consumption is normalized to the \texttt{interactive} CPU
  governor. Our approach gives the best trade-off.
  }\label{fig:pareto}
  \vspace{-2mm}
\end{figure}

%

\hl{In an attempt to explain the performance gains of our approach over eBrowser, we compare the distribution of the processor frequency
settings chosen by our approach, eBrowser and an Oracle approach. The Oracle approach is a theoretical perfect approach that always gives
the optimal processor setting. We determine the optimal processor setting by profiling all possible configurations on each webpage.}

\hl{Figure}~\ref{fig:conf_dis}  \hl{shows the results on each platform. The x-axis of the diagrams shows a processor configuration on a
hardware platform, and the y-axis shows how often a configuration is chosen in our evaluation dataset. The processor configurations are
sorted by their power consumption, from low to high. As can be seen from the diagram, eBrowser is in favor of high-power processor
configuration because it relies on the interactive CPU governor to control the CPU frequency. By contrast, our approach often chooses a
low-power processor configuration when possible. By exploiting web workload characteristics to actively exploiting the frequency control
knobs offered by the heterogeneous hardware design, our approach thus leads to better performance over eBrowser.}

Finally, Figure~\ref{fig:pareto} shows the Pareto efficiency of our approach, eBrowser, the \texttt{Interactive} and \texttt{Ondemand} CPU
governor\footnote{While \texttt{Ondemand} favours energy savings but it leads to significant QoS violations and is rarely used for
interactive applications.} when processing the landing page of \texttt{cnn.com} (see Section~\ref {sec:motivation}). From the diagram, we
see that our approach gives the best trade-off among all schemes for trading responsive time for energy reduction.

\subsection{FPS Prediction Accuracy}
\label{sec:predaccuracy}
\begin{figure}[t!]
  \centering
  \includegraphics[width=0.5\textwidth]{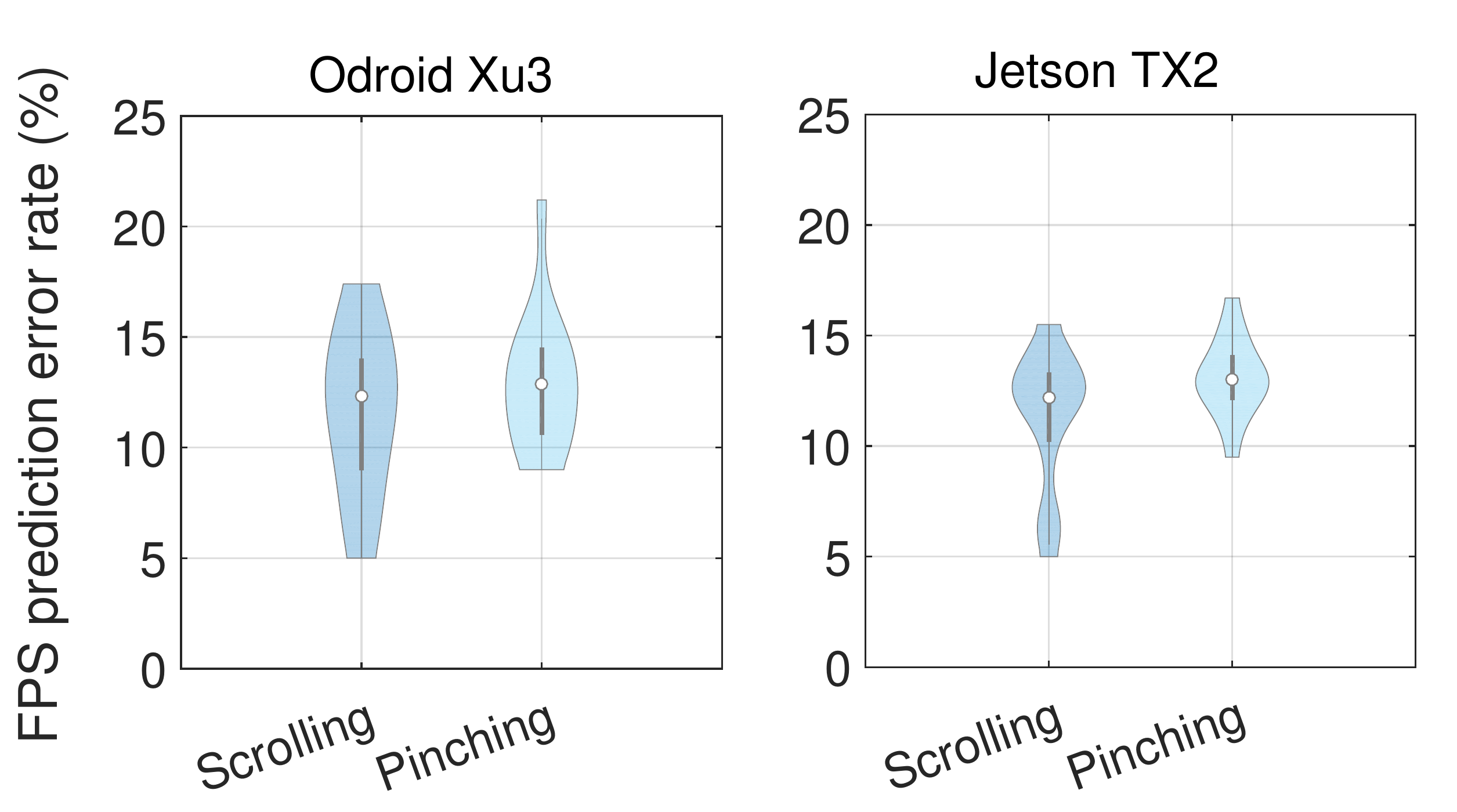}\\
  \caption{The FPS prediction errors. The thick line shows where 50\% of the data lines and the white dot shows the median value. Our approach gives a low prediction error of less than 15\% on both platforms. }\label{fig:errorrate}
\end{figure}

The violin plots in Figure~\ref{fig:errorrate} show the error rate for FPS value prediction for scrolling and pinching under the most
frequently used processor setting of each platform. The error, $e$, is calculated as:

$$e = \frac{|FPS_{measured} - FPS_{pred}|}{FPS_{measured}}
$$
where $FPS_{measured}$ and $FPS_{pred}$ are the measured and predicted FPS respectively.

In the diagram, the thick line shows where 50\% of the data lines. The white dot is the position of the median. Our predictive models are
highly accurate in predicting the FPS, with a mean error of less than 15\% on both evaluation platforms. The prediction accuracy can be
further improved by providing to the learning algorithm more training data, which also permits the use of a larger number of features to
better capture the application behavior. Nonetheless, our approach can give good results using as few as 80 training webpages.

\subsection{Processor Setting Distribution}
\begin{figure}
  \centering
  \subfloat[][Odroid Xu3]{\includegraphics[width=0.5\textwidth]{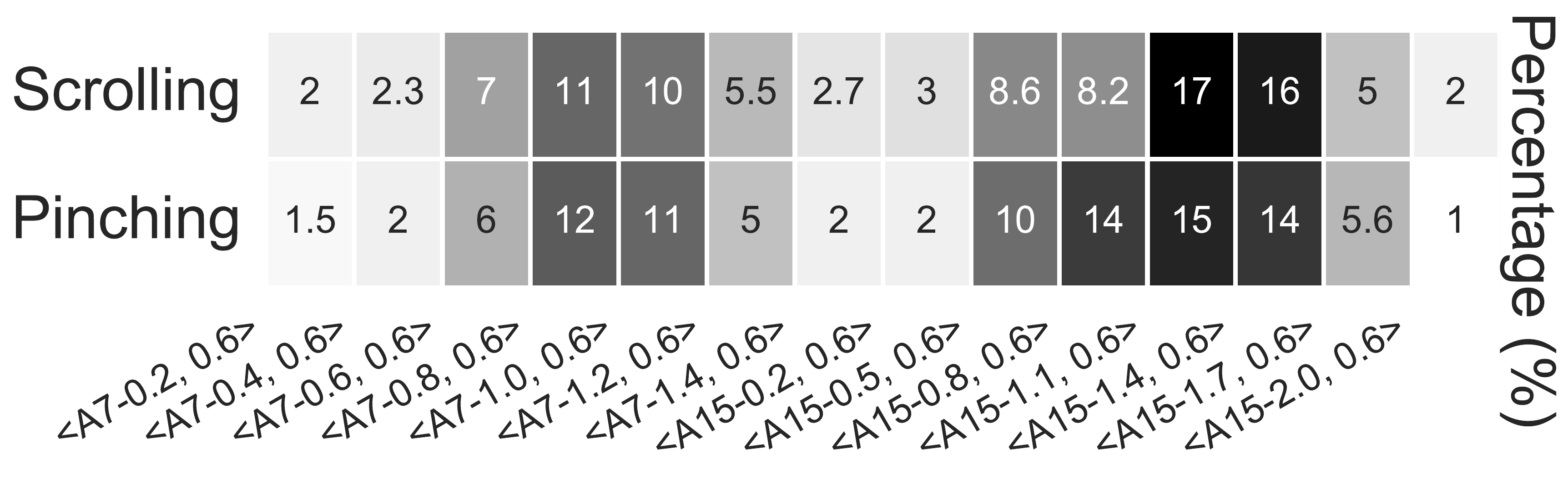}}\\
  \subfloat[][Jetson TX2]{\includegraphics[width=0.5\textwidth]{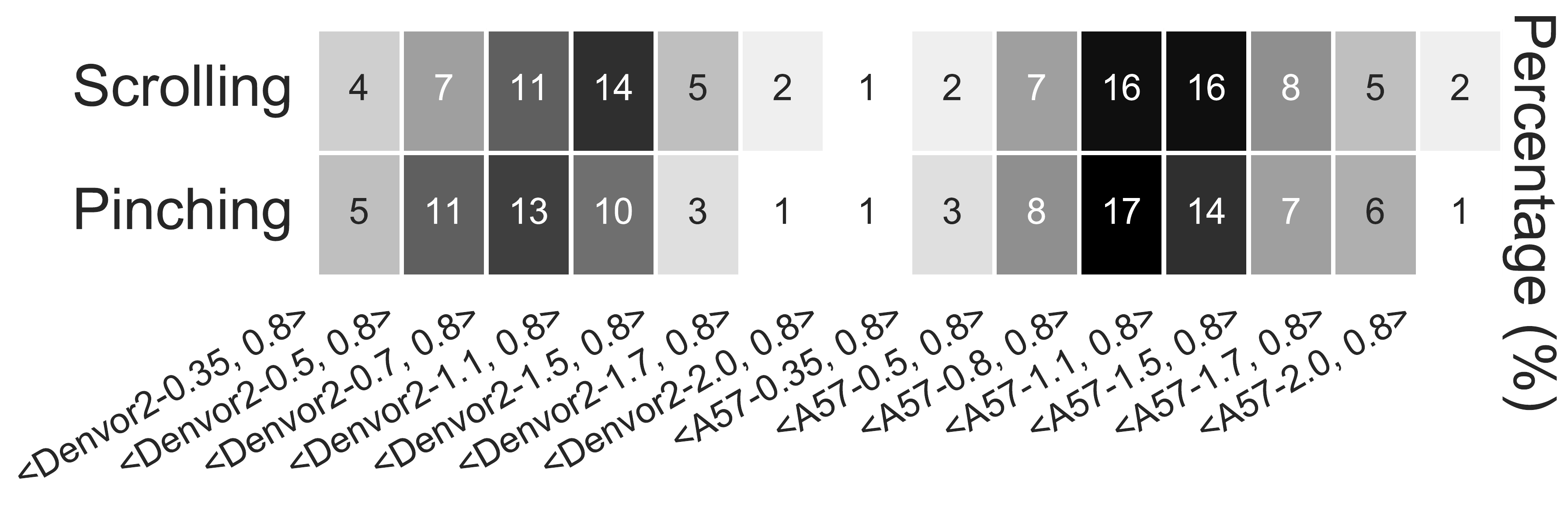}}
  \caption{How often (as percentages) a processor configuration is considered to be optimal by our model. There is no single configuration
  that is considered to be optimal for more than 20\% of the testing scenarios, suggesting the need for an adaptive scheme.
  }\label{fig:settingheat}
\end{figure}

The heat maps in Figure~\ref{fig:settingheat} depict how frequent a processor setting is chosen for pinching and scrolling on each of our
evaluation platforms. In the diagram, we use the notation $<$\emph{Rendering CPU cluster}-\emph{frequency}, \emph{frequency of the other
CPU cluster}$>$ to denote a processor configuration. For example, a configuration of $<A7-0.2, 0.6>$ on Odroid Xu3 means that the render
process running on the A7 core (little cluster) at 200Mhz and the remaining processes run on the A15 core (big cluster) at 600MHz;
similarly, a configuration of $<A57-1.1, 0.8>$ on Jetson TX2 means that the render process runs on the A57 core (big cluster) at 1.1GHz and
the remaining processes run on the Denvor2 core (little cluster) at 800MHz.

As can be seen from the diagram, there is no single processor configuration is considered to be optimal for more than 20\% of our testing
scenarios, and the frequency for a configuration to be optimal varies across hardware platforms. The results reinforce our claim that a
single ``one-size-fits-all" model is unlikely to deliver good performance across hardware architectures. Our work avoids this drawback by
developing a portable approach using machine learning.

\subsection{Feature Importance}
\label{sec:importance} 
\begin{figure}
  \centering
  \includegraphics[width=0.5\textwidth]{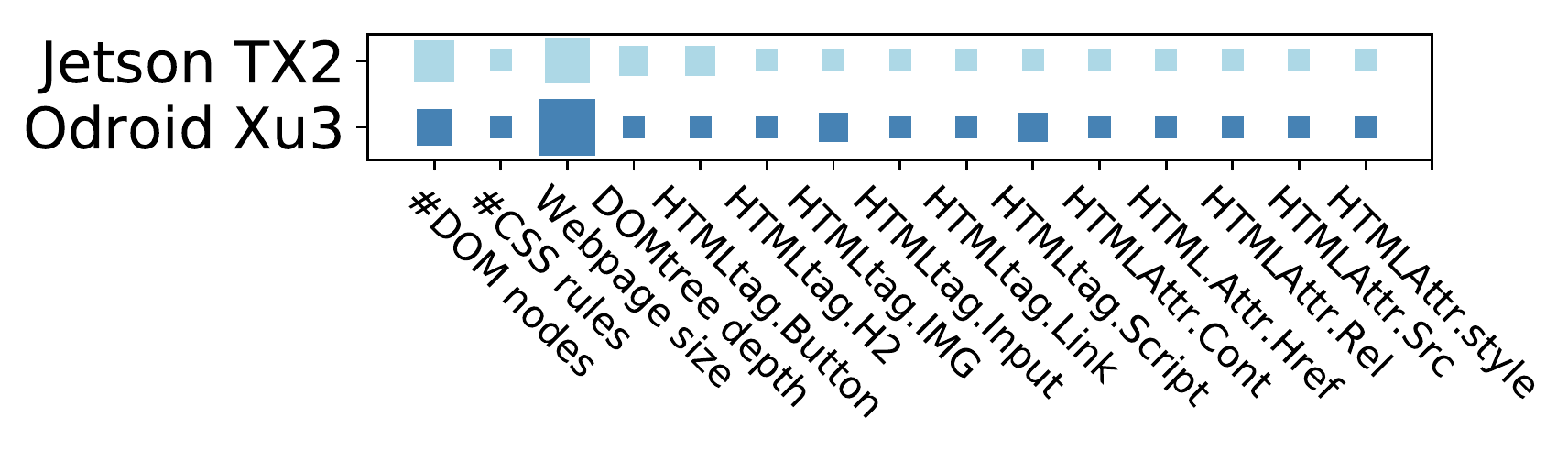}\\
  \caption{The Hinton diagram illustrates the importance of selected features for FPS predictions. The feature importance can vary across platforms, suggesting the need for an adaptive scheme.}\label{fig:featuresHinton}

\end{figure}

In an attempt to visualize what features are important for predicting the FPS, we plot a Hinton diagram in
Figure~\ref{fig:featuresHinton}. In the diagram, the larger the box, the more significantly a particular feature contributes to the
prediction accuracy on a given platform. The importance is calculated through the information gain ratio. It can be observed that HTML tags
and attributes (e.g. webpage size, \#DOM nodes, DOM tree depth) and style rules are useful when determining the processor configurations on
both platforms, but the importance varies across hardware architectures. We also observe that some features, like HTML tag.IMG and HTML
tag.Script, are useful for Odroid XU3 and are less important for Jetson TX2, which because Odroid Xu3 takes longer to process images and
JavaScript over Jetson TX due to its less powerful computation capability. This diagram suggests a generic, platform-independent
optimization model~\cite{pes} is unlike to be effective across a diverse set of architectures.

\begin{figure}[t!]
  \centering
  \includegraphics[width=0.45\textwidth]{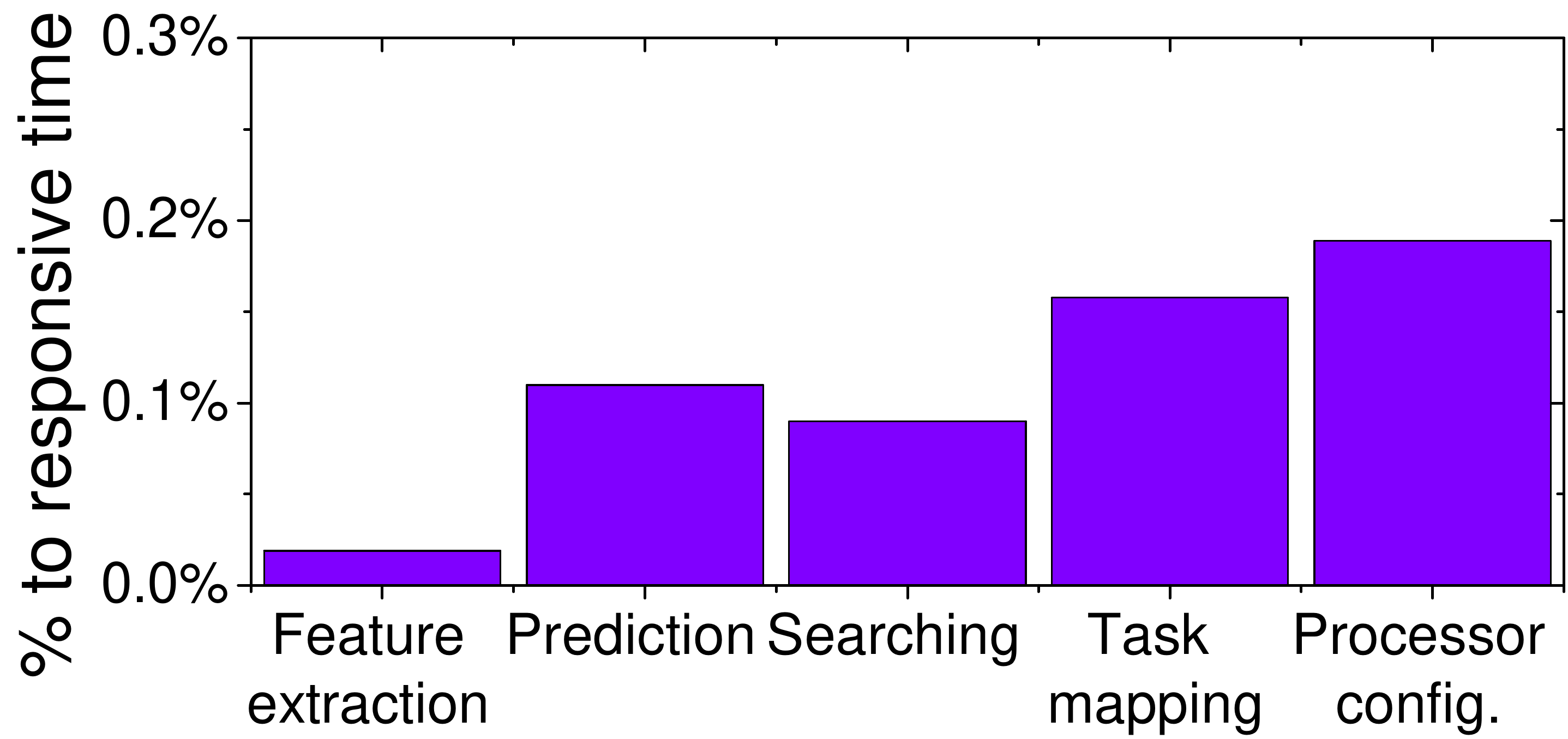}\\
  \caption{Breakdown of runtime overhead. Our approach incurs little runtime
overhead. }\label{fig:overhead}
\end{figure}

\subsection{Overhead Breakdown}
Figure~\ref{fig:overhead} gives a breakdown of the runtime overhead of our approach (which was already included in our experimental results).
The overhead of our approach including feature extraction,  prediction and searching, and task mapping and processor configuration. Feature
extraction typically only needs to perform once after the DOM tree has been constructed. Task migration and processor frequency setting
account for the majority of the overhead, but is less than 0.3\% of the end-to-end turnaround time. Such a small overhead can be easily
amortized by improved energy efficiency. We note that the user does not experience the training overhead as training data generation and
learning were performed \emph{off-line}.

\subsection{Alternative Predictive Modeling Techniques}
\label{sec:alter} 
\begin{figure}
  \centering
  \includegraphics[width=0.42\textwidth]{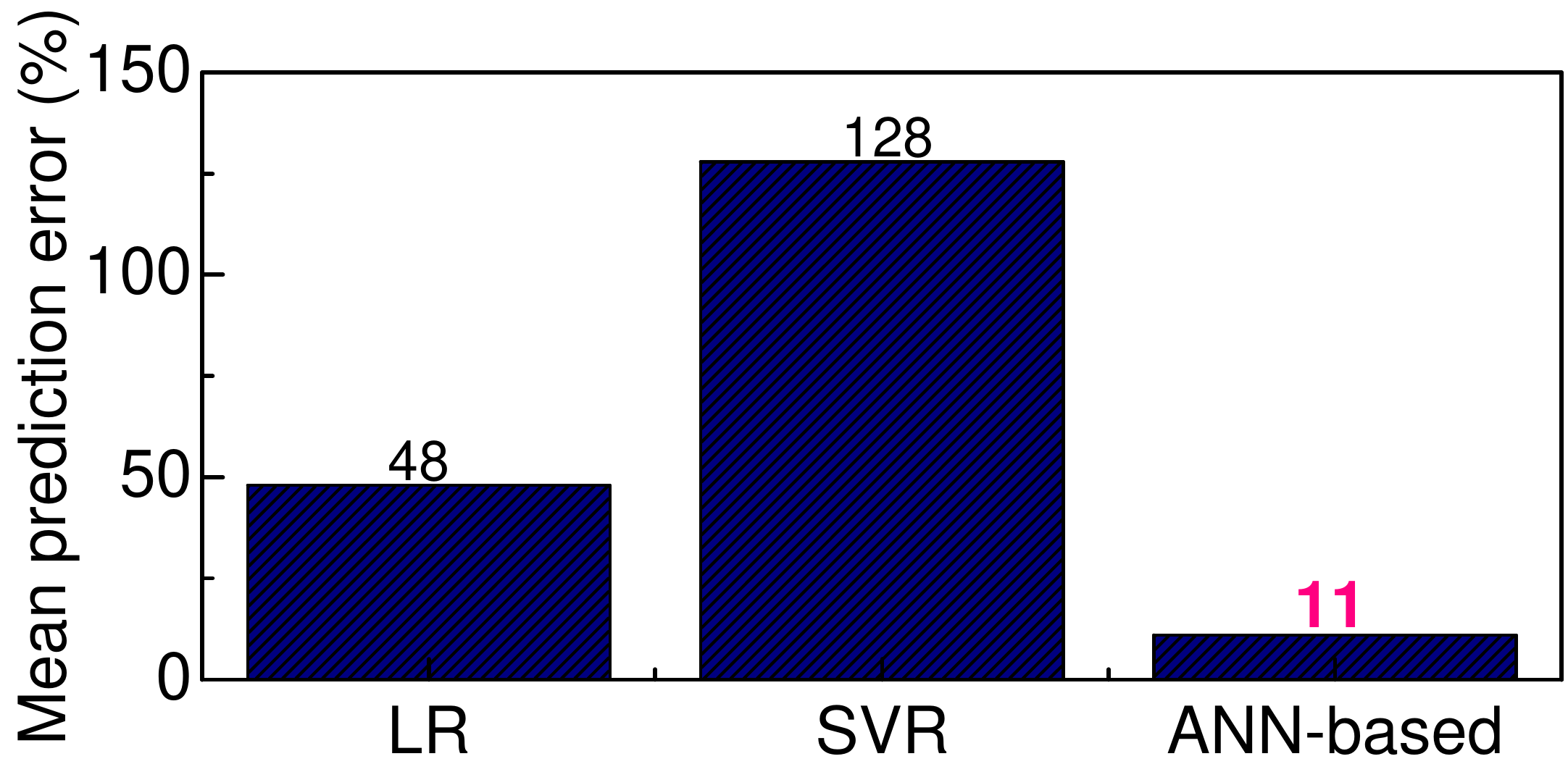}\\
  \caption{The mean FPS prediction error of our ANN-based approach, LR and SVR. Our approach gives the lowest mean prediction error. }\label{fig:modelCompare}
\end{figure}

We compare our ANN-based FPS predictor and two widely used regression techniques: linear regression (LR) and support
vector regression (SVR). For a fair comparison, we train and evaluate all techniques on the same dataset. Figure~\ref{fig:modelCompare}
shows the mean prediction error given by each modeling technique. Our approach gives the most accurate prediction results with the least
mean error across testing web pages, which is 77\% and 91\% lower than the LR and SVR counterparts respectively.

\begin{figure}
  \centering
  \includegraphics[width=0.42\textwidth]{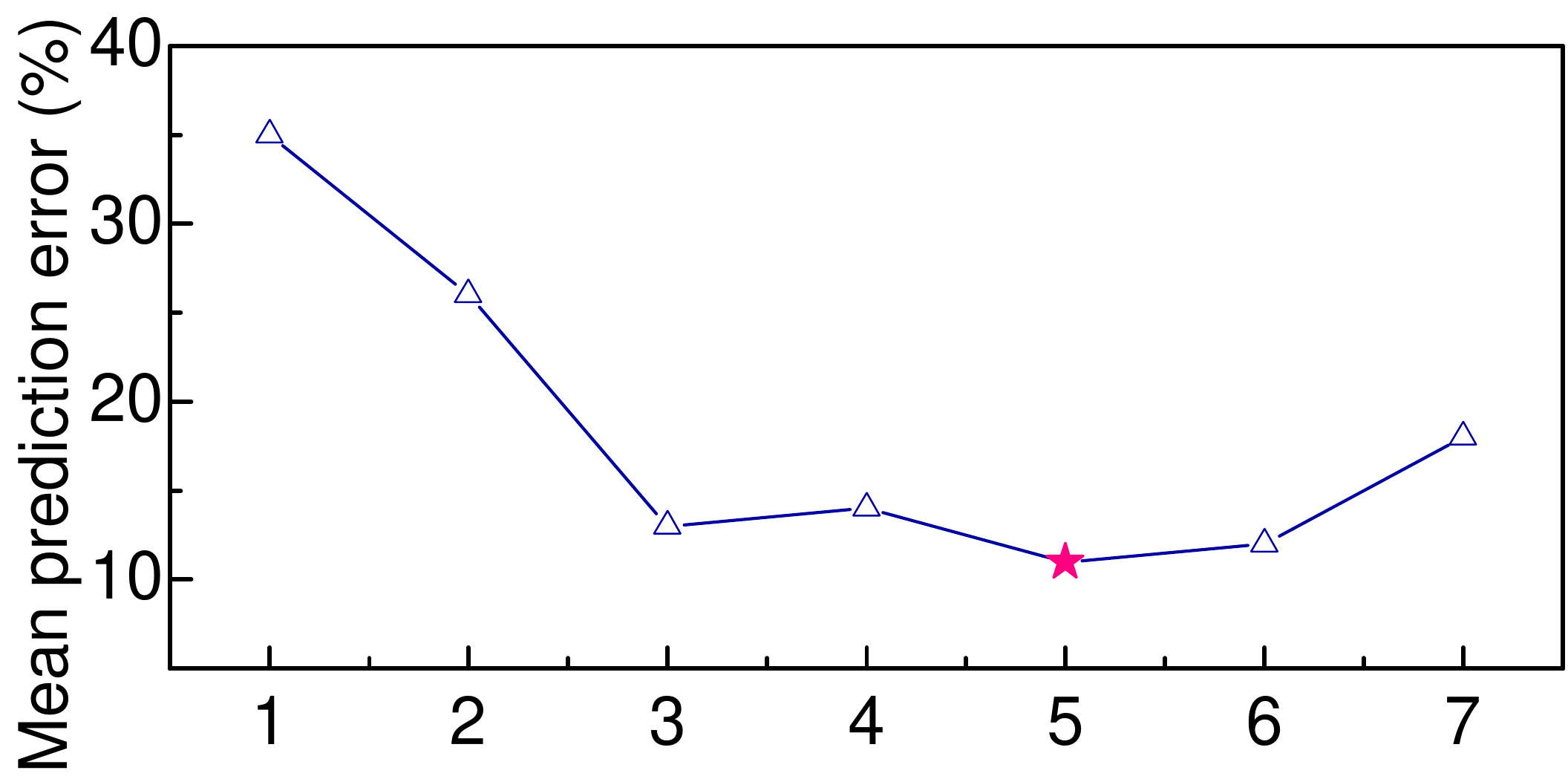}\\
  \caption{Mean FPS prediction errors with different number of neural hidden layers.}\label{fig:diffLayers}
\end{figure}

Figure~\ref{fig:diffLayers} shows how the FPS prediction error changes when different numbers of hidden layers are used for our ANN model.
Increasing the number of layers leads to a slightly improved prediction accuracy, but it reaches a plateau after five layers. Using more
than five layers would lead to a drop in accuracy, which is mainly attributed to our relatively small training dataset. In this paper, we
choose an ANN of five layers as it gives the smallest prediction error and does not require a large training dataset to learn.

\section{Discussions and Future Work}
Our work represents a new attempt for energy-efficient mobile web interactions through the use of machine learning. Like many research
works, there is room for further work and improvements. In this section, we discuss a few points on how we can improve the work.

\cparagraph{Performance Portability} A practical problem of a supervised learning approach is how to ensure the model can target a wide
range of devices and users. While retraining the model using data collected from the deployment environment can help to re-target an
existing model to a new environment, this can incur significant overhead for training data collection. Our future work will investigate how
to quickly port a decision model to a new environment.

\cparagraph{Multi-tasking mobile workloads} Our work assumes the user is interacting with one webpage at a time. This is a reasonable
assumption for mobile applications as unlike desktop PCs, there is typically only one foreground task which the user is dealing with;
background programs on mobile devices are typically put into a suspended (sleeping) or closed status. Nonetheless, our approach can be
extended to a multi-tasking computing environment that consists of multiple concurrently running workloads. This can be achieved by
triggering our scheduler when a web view is presented to the user.

\cparagraph{Network latency} Our work focuses on the interactive stage after a page has been loaded and processed to construct the DOM
tree. Hence, we do not consider the impact of network latency. It is possible that a user interaction might trigger a new download
activity, e.g., loading a new image. This is not explicitly modeled by our approach. However, there is work on energy optimization for page
loading, which considers the impact of networks~\cite{Ren:2018:PNW:3281411.3281422}. Such work is orthogonal to our approach. Extending our
work to consider the impact of network latency during user interactions is our future work.

\cparagraph{Impact of GPU frequency settings} Our work does not model the impact of GPU frequency. Instead, we rely on the default GPU
frequency governor to do so. However, our approach can be extended to dynamically adjust the GPU frequency. This would require us to
collect empirical data to learn how the GPU frequency setting affect the FPS. Training data collection and learning can be performed
automatically in the same way as we did throughout the work, and our methodology for model training and deployment can remain unchanged.
We leave this as our future work.

\cparagraph{Dynamic content} Our techniques were evaluated on static web content primarily consist of HTML files and images, which remain
the dominant content for mobile web applications. To target dynamic content such as JavaScript content or video streaming, we will need new
features to capture the workloads and a mechanism for constant monitoring and frequency adjustment. Given the dynamic nature of the
problem, it might be interesting to investigate whether a reinforcement learning based approach~\cite{sutton1998introduction} can be better
capture the behavior of the application domain.

\cparagraph{Impact of displays} Our work does not leverage the correlation between the web content and on-screen displays to further reduce
energy consumption. Nonetheless, our approach can be easily integrated with a display-based energy optimization scheme (which utilizing the
color and brightness settings to save energy), as the processor setting, in general, is independent on the colors and brightness of the
screen. We investigate such a holistic approach in our future work.

\section{Conclusions}
\hl{This paper has presented a novel machine learning based approach for optimizing interactive mobile web browsing.} At the heart of our
approach is a set of machine-learning-based regression models for predicting the resultant FPS under a given task to core and processor
frequency setting. The predictive models are first trained offline using training web pages and then used at runtime as a cost function to
quickly search for the optimal processor setting for new, unseen web content. We demonstrate that by carefully trading the responsive time,
one can significantly reduce the energy consumption during the interaction phase of mobile web browsing. We show that such energy reduction
can be achieved without significantly compromising the user-perceived latency or QoS.

We apply our approach to two representative mobile interactive events. We implement our methods in the open-source
 Chromium web browser, and thoroughly evaluated the developed system on two distinct heterogeneous mobile platforms using the landing
pages of top-100 popular websites. Experimental results show that our approach outperforms the state-of-the-art across webpages and
evaluation platforms and criteria. On average, our approach reduces the energy consumption by over 17\% over the state-of-the-art, and it
achieves this with fewer QoS violations.

\bibliographystyle{unsrt}

\bibliography{refs,zheng}

\begin{thebibliography}{10}

\bibitem{marketreport}
Dave Chaffey.
\newblock Mobile marketing statistics compilation.
\newblock
  \url{https://www.smartinsights.com/mobile-marketing/mobile-marketing-analytics/mobile-marketing-statistics/},
  2018.

\bibitem{mobiletraffic}
Greg Sterling.
\newblock Morgan stanley: No, apps aren't winning. the mobile browser is.
\newblock
  \url{https://marketingland.com/morgan-stanley-no-apps-arent-winning-the-mobile-browser-is-144303},
  2015.

\bibitem{ren2017optimise}
Jie Ren et~al.
\newblock Optimise web browsing on heterogeneous mobile platforms: a machine
  learning based approach.
\newblock In {\em INFOCOM '17}, 2017.

\bibitem{Ren:2018:PNW:3281411.3281422}
Jie Ren, Xiaoming Wang, Jianbin Fang, Yansong Feng, Dongxiao Zhu, Zhunchen Luo,
  Jie Zheng, and Zheng Wang.
\newblock Proteus: Network-aware web browsing on heterogeneous mobile systems.
\newblock In {\em Proceedings of the 14th International Conference on Emerging
  Networking EXperiments and Technologies}, CoNEXT '18, 2018.

\bibitem{pes}
Yu~Feng and Yuhao Zhu.
\newblock Pes: Proactive event scheduling for responsive and energy-efficient
  mobile web computing.
\newblock In {\em International Symposium on Computer Architecture (ISCA)},
  2019.

\bibitem{xu2018ebrowser}
Fei Xu, Shuai Yang, Zhi Zhou, and Jia Rao.
\newblock ebrowser: Making human-mobile web interactions energy efficient with
  event rate learning.
\newblock In {\em 2018 IEEE 38th International Conference on Distributed
  Computing Systems (ICDCS)}, pages 523--533, 2018.

\bibitem{wang2018machine}
Zheng Wang and Michael O'Boyle.
\newblock Machine learning in compiler optimization.
\newblock {\em Proceedings of the IEEE}, (99):1--23, 2018.

\bibitem{biglittle}
ARM ltd.
\newblock Arm big.little.
\newblock \url{https://www.arm.com/why-arm/technologies/big-little}.

\bibitem{seeker2014measuring}
Volker Seeker, Pavlos Petoumenos, Hugh Leather, and Bj{\"o}rn Franke.
\newblock Measuring qoe of interactive workloads and characterising frequency
  governors on mobile devices.
\newblock In {\em 2014 IEEE International Symposium on Workload
  Characterization (IISWC)}, pages 61--70. IEEE, 2014.

\bibitem{Chromium}
The chromium projects.
\newblock \url{https://www.chromium.org/}, 2019.

\bibitem{alexa}
Alexa.
\newblock \url{http://www.alexa.com/topsites}, 2019.

\bibitem{wang2015automatic}
Zheng Wang, Dominik Grewe, and Michael~FP O'boyle.
\newblock Automatic and portable mapping of data parallel programs to opencl
  for gpu-based heterogeneous systems.
\newblock {\em ACM Transactions on Architecture and Code Optimization (TACO)},
  11(4):42, 2015.

\bibitem{moran2018overcoming}
Kevin Moran, Carlos Bernal-C{\'a}rdenas, Mario Linares-V{\'a}squez, and Denys
  Poshyvanyk.
\newblock Overcoming language dichotomies: toward effective program
  comprehension for mobile app development.
\newblock In {\em Proceedings of the 26th Conference on Program Comprehension},
  pages 7--18. ACM, 2018.

\bibitem{Taylor:2017:AOO:3078633.3081040}
Ben Taylor, Vicent~Sanz Marco, and Zheng Wang.
\newblock Adaptive optimization for opencl programs on embedded heterogeneous
  systems.
\newblock In {\em Proceedings of the 18th ACM SIGPLAN/SIGBED Conference on
  Languages, Compilers, and Tools for Embedded Systems}, LCTES 2017, 2017.

\bibitem{taylor2018adaptive}
Ben Taylor, Vicent~Sanz Marco, Willy Wolff, Yehia Elkhatib, and Zheng Wang.
\newblock Adaptive deep learning model selection on embedded systems.
\newblock In {\em Proceedings of the 19th ACM SIGPLAN/SIGBED International
  Conference on Languages, Compilers, and Tools for Embedded Systems}, pages
  31--43. ACM, 2018.

\bibitem{6848020}
W.~{Hu} and G.~{Cao}.
\newblock Energy optimization through traffic aggregation in wireless networks.
\newblock In {\em IEEE INFOCOM 2014 - IEEE Conference on Computer
  Communications}, pages 916--924, 2014.

\bibitem{li2016automated}
Ding Li, Yingjun Lyu, Jiaping Gui, and William~GJ Halfond.
\newblock Automated energy optimization of http requests for mobile
  applications.
\newblock In {\em Proceedings of the 38th international conference on software
  engineering}, pages 249--260. ACM, 2016.

\bibitem{sehati2017energy}
Ali Sehati and Majid Ghaderi.
\newblock Energy-delay tradeoff for request bundling on smartphones.
\newblock In {\em IEEE INFOCOM 2017-IEEE Conference on Computer
  Communications}, pages 1--9. IEEE, 2017.

\bibitem{zhu2013high}
Yuhao Zhu and Vijay~Janapa Reddi.
\newblock High-performance and energy-efficient mobile web browsing on
  big/little systems.
\newblock In {\em 2013 IEEE 19th International Symposium on High Performance
  Computer Architecture (HPCA)}, pages 13--24. IEEE, 2013.

\bibitem{Peters:2018:PWB:3205289.3205293}
N.~Peters, S.~Park, D.~Clifford, S.~Kyostila, R.~Mcllroy, B.~Meurer, H.~Payer,
  and S.~Chakraborty.
\newblock Phase-aware web browser power management on hmp platforms.
\newblock In {\em Proceedings of the 2018 International Conference on
  Supercomputing}, ICS '18, 2018.

\bibitem{thiagarajan2012killed}
Narendran Thiagarajan, Gaurav Aggarwal, Angela Nicoara, Dan Boneh, and
  Jatinder~Pal Singh.
\newblock Who killed my battery?: analyzing mobile browser energy consumption.
\newblock In {\em Proceedings of the 21st international conference on World
  Wide Web}, pages 41--50. ACM, 2012.

\bibitem{cao2017deconstructing}
Yi~Cao, Javad Nejati, Muhammad Wajahat, Aruna Balasubramanian, and Anshul
  Gandhi.
\newblock Deconstructing the energy consumption of the mobile page load.
\newblock {\em Proceedings of the ACM on Measurement and Analysis of Computing
  Systems}, 1(1):6, 2017.

\bibitem{butkiewicz2015klotski}
Michael Butkiewicz, Daimeng Wang, Zhe Wu, Harsha~V Madhyastha, and Vyas Sekar.
\newblock Klotski: Reprioritizing web content to improve user experience on
  mobile devices.
\newblock In {\em 12th $\{$USENIX$\}$ Symposium on Networked Systems Design and
  Implementation ($\{$NSDI$\}$ 15)}, pages 439--453, 2015.

\bibitem{netravali2018prophecy}
Ravi Netravali and James Mickens.
\newblock Prophecy: accelerating mobile page loads using final-state write
  logs.
\newblock In {\em 15th $\{$USENIX$\}$ Symposium on Networked Systems Design and
  Implementation ($\{$NSDI$\}$ 18)}, pages 249--266, 2018.

\bibitem{Qian:2014:CRU:2594368.2594372}
Feng Qian, Subhabrata Sen, and Oliver Spatscheck.
\newblock Characterizing resource usage for mobile web browsing.
\newblock In {\em Proceedings of the 12th Annual International Conference on
  Mobile Systems, Applications, and Services}, MobiSys '14, 2014.

\bibitem{bui2015rethinking}
Duc~Hoang Bui, Yunxin Liu, Hyosu Kim, Insik Shin, and Feng Zhao.
\newblock Rethinking energy-performance trade-off in mobile web page loading.
\newblock In {\em Proceedings of the 21st Annual International Conference on
  Mobile Computing and Networking}, pages 14--26. ACM, 2015.

\bibitem{Augonnet:2011:SUP:1951453.1951454}
CEdric Augonnet and OTHERS.
\newblock Starpu: A unified platform for task scheduling on heterogeneous
  multicore architectures.
\newblock {\em Concurr. Comput. : Pract. Exper.}, 2011.

\bibitem{mittal2015survey}
Sparsh Mittal and Jeffrey~S Vetter.
\newblock A survey of cpu-gpu heterogeneous computing techniques.
\newblock {\em ACM Computing Surveys (CSUR)}, 47(4):69, 2015.

\bibitem{chronaki2016task}
Kallia Chronaki, Alejandro Rico, Marc Casas, Miquel Moret{\'o}, Rosa~M Badia,
  Eduard Ayguad{\'e}, Jesus Labarta, and Mateo Valero.
\newblock Task scheduling techniques for asymmetric multi-core systems.
\newblock {\em IEEE Transactions on Parallel and Distributed Systems},
  28(7):2074--2087, 2016.

\bibitem{castillo2018architectural}
Emilio Castillo, Lluc Alvarez, Miquel Moreto, Marc Casas, Enrique Vallejo,
  Jose~Luis Bosque, Ramon Beivide, and Mateo Valero.
\newblock Architectural support for task dependence management with flexible
  software scheduling.
\newblock In {\em 2018 IEEE International Symposium on High Performance
  Computer Architecture (HPCA)}, pages 283--295. IEEE, 2018.

\bibitem{krause2005context}
Andreas Krause, Asim Smailagic, and Daniel~P Siewiorek.
\newblock Context-aware mobile computing: Learning context-dependent personal
  preferences from a wearable sensor array.
\newblock {\em IEEE Transactions on Mobile Computing}, 5(2):113--127, 2005.

\bibitem{plotz2018deep}
Thomas Pl{\"o}tz and Yu~Guan.
\newblock Deep learning for human activity recognition in mobile computing.
\newblock {\em Computer}, 51(5):50--59, 2018.

\bibitem{Zhang:2018:CTC:3241539.3241570}
Jie Zhang, Zhanyong Tang, Meng Li, Dingyi Fang, Petteri Nurmi, and Zheng Wang.
\newblock Crosssense: Towards cross-site and large-scale wifi sensing.
\newblock In {\em Proceedings of the 24th Annual International Conference on
  Mobile Computing and Networking}, MobiCom '18, 2018.

\bibitem{wang2014integrating}
Zheng Wang et~al.
\newblock Integrating profile-driven parallelism detection and
  machine-learning-based mapping.
\newblock {\em ACM TACO}, 2014.

\bibitem{Tournavitis:2009:THA:1542476.1542496}
Georgios Tournavitis et~al.
\newblock Towards a holistic approach to auto-parallelization: Integrating
  profile-driven parallelism detection and machine-learning based mapping.
\newblock In {\em PLDI '09}, 2009.

\bibitem{Wang:2009:MPM:1504176.1504189}
Zheng Wang and Michael~F.P. O'Boyle.
\newblock Mapping parallelism to multi-cores: A machine learning based
  approach.
\newblock In {\em PPoPP '09}, 2009.

\bibitem{wang2010partitioning}
Zheng Wang and Michael~FP O'Boyle.
\newblock Partitioning streaming parallelism for multi-cores: a machine
  learning based approach.
\newblock In {\em PACT '10}, 2010.

\bibitem{grewe2013portable}
Dominik Grewe et~al.
\newblock Portable mapping of data parallel programs to opencl for
  heterogeneous systems.
\newblock In {\em CGO}, 2013.

\bibitem{wang2013using}
Zheng Wang and Michael~FP O'boyle.
\newblock Using machine learning to partition streaming programs.
\newblock {\em ACM TACO}, 2013.

\bibitem{DBLP:journals/taco/WangGO14}
Zheng Wang et~al.
\newblock Automatic and portable mapping of data parallel programs to opencl
  for gpu-based heterogeneous systems.
\newblock {\em {ACM TACO}}, 2014.

\bibitem{ogilvie2014fast}
William~F Ogilvie et~al.
\newblock Fast automatic heuristic construction using active learning.
\newblock In {\em LCPC '14}, 2014.

\bibitem{cummins2017end}
Chris Cummins et~al.
\newblock End-to-end deep learning of optimization heuristics.
\newblock In {\em PACT '17}, 2017.

\bibitem{ogilvie2017minimizing}
William~F Ogilvie et~al.
\newblock Minimizing the cost of iterative compilation with active learning.
\newblock In {\em CGO '17}, 2017.

\bibitem{spmv}
Shizhao Chen et~al.
\newblock Adaptive optimization of sparse matrix-vector multiplication on
  emerging many-core architectures.
\newblock In {\em HPCC '18}, 2018.

\bibitem{ipdpsz18}
Peng Zhang, , et~al.
\newblock Auto-tuning streamed applications on intel xeon phi.
\newblock In {\em IPDPS '18}, 2018.

\bibitem{ijpp18}
Chen Lindong et~al.
\newblock Optimizing sparse matrix-vector multiplications on an armv8-based
  many-core architecture.
\newblock {\em International Journal of Parallel Programming}, 2018.

\bibitem{grewe2011workload}
Dominik Grewe et~al.
\newblock A workload-aware mapping approach for data-parallel programs.
\newblock In {\em HiPEAC '11}, 2011.

\bibitem{emani2013smart}
Murali~Krishna Emani et~al.
\newblock Smart, adaptive mapping of parallelism in the presence of external
  workload.
\newblock In {\em CGO '13}, 2013.

\bibitem{grewe2013opencl}
Dominik Grewe et~al.
\newblock Opencl task partitioning in the presence of gpu contention.
\newblock In {\em LCPC '13}, 2013.

\bibitem{wen2014smart}
Yuan Wen et~al.
\newblock Smart multi-task scheduling for opencl programs on cpu/gpu
  heterogeneous platforms.
\newblock In {\em HiPC '14}, 2014.

\bibitem{lo2015prediction}
Daniel Lo et~al.
\newblock Prediction-guided performance-energy trade-off for interactive
  applications.
\newblock In {\em Proceedings of the 48th International Symposium on
  Microarchitecture}, pages 508--520, 2015.

\bibitem{gaudette2016improving}
Benjamin Gaudette et~al.
\newblock Improving smartphone user experience by balancing performance and
  energy with probabilistic qos guarantee.
\newblock In {\em 2016 IEEE International Symposium on High Performance
  Computer Architecture (HPCA)}, pages 52--63, 2016.

\bibitem{Mishra:2018:CLC:3173162.3173184}
Nikita Mishra et~al.
\newblock Caloree: Learning control for predictable latency and low energy.
\newblock In {\em Proceedings of the Twenty-Third International Conference on
  Architectural Support for Programming Languages and Operating Systems},
  ASPLOS '18, 2018.

\bibitem{Seo2015Big}
Wonik Seo et~al.
\newblock Big or little: A study of mobile interactive applications on an
  asymmetric multi-core platform.
\newblock In {\em IEEE International Symposium on Workload Characterization},
  2015.

\bibitem{arm}
big.little technology.
\newblock
  \url{http://www.arm.com/products/processors/technologies/biglittleprocessing}.

\bibitem{maas2013rectifier}
Andrew~L Maas et~al.
\newblock Rectifier nonlinearities improve neural network acoustic models.
\newblock In {\em Proc. icml}, volume~30, page~3, 2013.

\bibitem{dunteman1989principal}
George~H Dunteman.
\newblock {\em Principal components analysis}.
\newblock Number~69. 1989.

\bibitem{manly2016multivariate}
Bryan~FJ Manly and Jorge A~Navarro Alberto.
\newblock {\em Multivariate statistical methods: a primer}.
\newblock CRC Press, 2016.

\bibitem{kingma2014adam}
Diederik~P Kingma and Jimmy Ba.
\newblock Adam: A method for stochastic optimization.
\newblock {\em arXiv}, 2014.

\bibitem{marco2017improving}
Vicent~Sanz Marco et~al.
\newblock Improving spark application throughput via memory aware task
  co-location: a mixture of experts approach.
\newblock In {\em Proceedings of the 18th ACM/IFIP/USENIX Middleware
  Conference}, 2017.

\bibitem{wu2019machine}
Carole-Jean Wu, David Brooks, Kevin Chen, Douglas Chen, Sy~Choudhury, Marat
  Dukhan, Kim Hazelwood, Eldad Isaac, Yangqing Jia, Bill Jia, et~al.
\newblock Machine learning at facebook: Understanding inference at the edge.
\newblock In {\em 2019 IEEE International Symposium on High Performance
  Computer Architecture (HPCA)}, pages 331--344, 2019.

\bibitem{ertel1994definition}
Wolfgang Ertel.
\newblock On the definition of speedup.
\newblock In {\em International Conference on Parallel Architectures and
  Languages Europe}, 1994.

\bibitem{sutton1998introduction}
Richard~S Sutton, Andrew~G Barto, et~al.
\newblock {\em Introduction to reinforcement learning}, volume 135.
\newblock MIT press Cambridge, 1998.

\end{thebibliography}


\balance

\end{document}